\documentclass[prl,twocolumn,longbibliography,superscriptaddress,preprintnumbers]{revtex4-2}

\usepackage[colorlinks,bookmarks=true,citecolor=blue,linkcolor=blue,urlcolor=blue] {hyperref}
\usepackage{dcolumn,graphicx,amsfonts,amsthm,bm,color,appendix,float}
\usepackage{braket}
\usepackage[normalem]{ulem}
\usepackage[version=4]{mhchem}
\usepackage{siunitx}
\usepackage{amsmath}
\usepackage{amssymb}
\usepackage{CJKutf8}
\usepackage{bbm}
\usepackage{subfigure}
\usepackage{color}

\theoremstyle{definition}
\usepackage{enumitem}

\usepackage[dvipsnames]{xcolor}
\definecolor{pal0}{rgb}{0.8941, 0.102 , 0.1098}
\definecolor{pal1}{rgb}{0.2157, 0.4941, 0.7216}
\definecolor{pal2}{rgb}{0.302 , 0.6863, 0.2902}
\definecolor{pal3}{rgb}{0.5961, 0.3059, 0.6392}
\definecolor{pal4}{rgb}{1.    , 0.498 , 0.    }

\begin{document}
\title{Quantum Geometry Probed by Chiral Excitonic Optical Response of Chern Insulators}

\author{Wen-Xuan Qiu}
\affiliation{School of Physics and Technology, Wuhan University, Wuhan 430072, China}
\author{Fengcheng Wu}
\email{wufcheng@whu.edu.cn}
\affiliation{School of Physics and Technology, Wuhan University, Wuhan 430072, China}
\affiliation{Wuhan Institute of Quantum Technology, Wuhan 430206, China}

\begin{abstract}
We theoretically derive the sum rule for the negative first moment of the absorptive optical conductivity with excitonic effects and establish its connection to the quantum weight $K$ and Chern number $C$ of the ground state. Applying this framework, we investigate the excitonic optical response of the Chern insulator at hole filling factor $\nu=1$ in twisted bilayer MoTe$_2$. A single chiral exciton state, which selectively absorbs circularly polarized light of a specific handedness, dominates the optical sum rule. The chiral exciton state comprises two types of interlayer electron-hole transitions, which cancel out the total out-of-plane dipole moment.  The absorption spectrum shows nearly perfect magnetic circular dichroism, which can be attributed to the nearly saturated bound $K \ge |C|$ of the Chern insulator under study.  Our work illustrates the potential of using excitonic optical responses to probe quantum geometry encoded by $K$ and $C$ of Chern insulators in moir\'e superlattices.
\end{abstract}

\maketitle
\textit{Introduction.---}The quantum geometric tensor of Bloch states consists of  Berry curvature and quantum metric, which characterize, respectively, the phase and amplitude distances between nearby quantum states \cite{TormaPaivi2023}. Quantum geometry, by capturing the structure of Bloch wavefunctions \cite{Provost1980,MaYuQuan2010}, plays an important role in charge transport \cite{Thouless1982,NiuQian1985,Haldane1988,Vanderbilt2018,GaoYang2014,Sodemann2015,Ma2019,WangChong2021,DasKamal2023,KaplanDaniel2024,AnyuanGao2023,WangNaizhou2023}, optical responses \cite{Souza2000,RestaRaffaeleP2006,Onishi2024prx,OnishiQuantumweight2024,Komissarov2024,VermaNishchhal2024,SouzaIvoDichroic2008,Cook2017,deJuan2017,AhnJunyeong2020,HolderTobias2020,BhallaPankaj2022,AversaClaudio1995,Orenstein2021,Ahn2022,KruchkovSpectral2023,SamuelBeaulieu2024}, and many-body physics \cite{Roy2014Band,Peotta2015,Ledwith2020,WangJie2021,VMbohao2024,TormaPaivi2022,TianEvidence2023,YuNontrivial2024,Srivastava2015,ZhouJianhui2015,WuFengcheng2020Quantum,XieHongYi2023,VermaNishchhal_stiffness}. For an insulator, the integral of the Berry curvature over the Brillouin zone gives the Chern number $C$, which is a well-known topological invariant that measures the Hall conductivity in a unit of $e^2/h$ \cite{Thouless1982}.  This Chern number $C$ also governs magnetic circular dichroism (MCD)  via a sum rule for the optical Hall conductivity \cite{SouzaIvoDichroic2008}. By comparison, the Brillouin-zone integral of the quantum metric trace has been identified as the quantum weight $K$, which lacks topological invariance but is observable, as it is proportional to the negative first moment of longitudinal optical conductivity \cite{Souza2000,RestaRaffaeleP2006,Onishi2024prx}. The impact of the quantum geometry on optical responses has resulted in a topological bound on band gap and a prediction of perfect MCD when the trace inequality $K\ge |C|$ is saturated \cite{Onishi2024prx,DongJunkai2023}. Numerical investigations have explored these phenomena in systems of twisted bilayer MoTe$_2$ ($t$MoTe$_2$) \cite{Onishi2024prx, DongJunkai2023} and MnBi$_2$Te$_4$ thin films \cite{GhoshBarun2024}. On the other hand, optical responses are often subjected to excitonic effects due to electron-hole interactions, which can dramatically change the optical spectrum. The relationship between the quantum geometry of the ground state and the excitonic optical response is a fundamental problem that remains to be studied.

In this Letter, we derive the sum rule for the negative first moment of the absorptive optical conductivity, assuming a Slater determinant ground state while also considering excitonic effects in the excited states. This sum rule is expressed using the ground-state quantum geometric quantities $K$ and $C$, which establishes a way to probe quantum geometry using excitonic optical response. As an illustrative example, we investigate the optical response of the Chern insulator (CI), also known as the quantum anomalous Hall insulator, in $t$MoTe$_2$ at  $\nu=1$. Here $t$MoTe$_2$ represents a key moir\'e system as it hosts both integer and fractional Chern insulators at zero external magnetic field \cite{Xiaodong2023a,Xiaodong2023b,Yihang2023_integer,Park2023b,Xu2023}. It is of great scientific significance in probing the quantum geometry of Chern bands in order to examine how close the bound $K \ge |C|$ is saturated. The motivation is that the saturated bound $K = |C|$ signals the resemblance of a Chern band and the lowest Landau level, which has been proposed as one of the ideal conditions for the realization of fractional Chern insulators \cite{Ledwith2020,WangJie2021,VMbohao2024}.
    
We calculate the exciton states on top of the CI ground state in $t$MoTe$_2$ at $\nu=1$ by solving the Bethe-Salpeter equation. The optical spectrum exhibits pronounced excitonic effects, featuring a single dominant chiral exciton state. We numerically validate the derived optical sum rule and find a significant contribution from the exciton state. The system exhibits a contrasting difference in the absorption of left and right circularly polarized light with nearly perfect MCD, which reflects the nearly saturated bound $K \ge |C|$ of the ground state.  We further elucidate that while optical sum rules can bound the optical gap, they do not constrain the charge gap. Additionally, we conduct a comparative investigation of the optical spectrum in a topologically trivial state. Our study underscores the important role of quantum geometry in constraining excitonic optical response and lays a foundation for optically probing the quantum geometry.

\textit{Optical sum rules.---}The optical conductivity $\sigma_{\alpha \beta}$ for an insulator as a function of frequency $\omega$ at zero temperature is given by Kubo formula, 
\begin{equation}\label{kubo}
\sigma_{\alpha \beta}(\omega)=i \frac{e^2}{\hbar} \frac{1}{\mathcal{A}} \sum_\chi \frac{1}{\mathcal{E}_\chi}\left[\frac{V_{G \chi}^\alpha V_{\chi G}^\beta}{\hbar \omega-\mathcal{E}_\chi+i \eta}+\frac{V_{\chi G}^\alpha V_{G \chi}^\beta}{\hbar \omega+\mathcal{E}_\chi+i \eta}\right],
\end{equation}
where $e$ is the elementary charge, $\mathcal{A}$ is the system size, $V_{\chi G}^\alpha=\langle \chi|\hat{v}^{\alpha}|G \rangle$ is the optical matrix element with velocity operator $\hat{\bm v}=i[\hat{\mathcal{H}},\bm \hat{\bm r}]$ along direction $\alpha$. Here $\bm \hat{\bm r}$ is the position operator of a many-body system [see definition in Eq.~\eqref{roperator}], $\hat{\mathcal{H}}$ is the Hamiltonian, $|G\rangle$ is the ground state, $\mathcal{E}_\chi$ is the energy of an excited state $|\chi \rangle$ measured relative to that of $|G\rangle$, and $\eta \rightarrow 0^+$. We focus on two-dimensional systems with an out-of-plane threefold rotational symmetry $\hat{C}_{3z}$, where $\sigma_{xx}=\sigma_{yy}$ and $\sigma_{xy}=-\sigma_{yx}$. 
The absorptive part of optical conductivity can then be defined as  $\sigma^{\text{abs}}_{\alpha\beta}=\delta_{\alpha\beta}\text{Re}\sigma_{{\alpha\beta}}+i(1-\delta_{\alpha\beta})\text{Im}\sigma_{{\alpha\beta}}$. The negative first  moment $W^{(1)}_{\alpha\beta}$ of $\sigma^{\text{abs}}_{\alpha\beta}$ is calculated to be, 
\begin{equation} \label{fnlr1}
\begin{aligned}
W^{(1)}_{\alpha\beta}=&\int_0^{\infty} d \omega \frac{\sigma_{\alpha \beta}^{\mathrm{abs}}(\omega)}{\omega}=\frac{e^2}{\hbar} \frac{\pi}{\mathcal{A}} \sum_\chi w^{(1)}_{\alpha\beta}(\chi) \\
=&\frac{e^2}{\hbar} \frac{\pi}{\mathcal{A}}[\left\langle \hat{r}^\alpha \hat{r}^\beta\right\rangle_{G}-\left\langle \hat{r}^\alpha\right\rangle_{G}\left\langle \hat{r}^\beta\right\rangle_{G}], 
\end{aligned}
\end{equation}
where $w^{(1)}_{\alpha\beta}(\chi)=V_{G \chi}^\alpha V_{\chi G}^\beta/\mathcal{E}_\chi^2$ and $\left\langle\cdots\right\rangle_G$ represents the ground state expectation value. Equation~\eqref{fnlr1} connects the generalized optical weight $W^{(1)}_{\alpha \beta}$ with the mean-square fluctuation of polarization in the ground state \cite{Souza2000}. $W^{(1)}_{\alpha \beta}$ can be further expressed using the quantum geometric quantities of the many-body ground state under twisted boundary conditions\cite{Souza2000,RestaRaffaeleP2006,Onishi2024prx}.

To avoid the complexity involved in the twisted boundary condition while making analytical progress, we focus on the special yet widely applicable case where $|G \rangle$ can be approximated by a Slater determinant composed of occupied Bloch states $|o \bm k \rangle$. Here $o$ is the band index and $\bm k$ labels momentum. The unoccupied  Bloch states are denoted as $|\bar{o} \bm k \rangle$. In the complete Bloch basis, the position operator $\bm \hat{\bm r}$ in the second quantized form is \cite{Karplus1954,Ahn2022},
\begin{equation}
\begin{aligned}
&\hat{\bm r}=\sum_{\bm k} \sum_{mn}
\bm r_{mn}(\bm k) f_{\bm k,m}^{\dagger} f_{\bm k,n}, \\
&\bm r_{mn}(\bm k)=\delta_{mn} i \partial_{\bm k}-\bm A_{mn}(\bm k),\\
&\bm A_{mn}(\bm k)=-\left\langle u_{m \bm k}\left|i \partial_{\bm k}\right| u_{n \bm k}\right\rangle,
\end{aligned}
\label{roperator}
\end{equation}
where $|u_{n \bm k}\rangle$ is the periodic part of the Bloch state $|n \bm k\rangle $ and $\bm A_{mn}(\bm k)$ is the Berry connection. We take Eq.~\eqref{roperator} as the operative definition of $\hat{\bm r}$.  $W_{\alpha \beta}^{(1)}$ is then expressed as,
\begin{equation}\label{fnlr2}
\begin{aligned}
W_{\alpha \beta}^{(1)}
=\frac{e^2}{\hbar} \frac{\pi}{\mathcal{A}}\sum_{\bm k}\sum_{\bar{o}o} A_{o \bar{o}}^\alpha(\bm k) A_{\bar{o} o}^\beta (\bm k) 
=\frac{e^2}{\hbar} \frac{\pi}{\mathcal{A}}\sum_{\bm k}\text{Tr}\mathcal{Q}^{\alpha\beta}(\bm k),
\end{aligned}
\end{equation}
where $\mathcal{Q}^{\alpha\beta}_{oo'}=\sum_{\bar{o}} A_{o \bar{o}}^\alpha A_{\bar{o} o'}^\beta$ is the quantum geometric tensor of occupied bands \cite{MaYuQuan2010}  and  $(\mathcal{Q}^{\alpha\beta})^{\dagger}=\mathcal{Q}^{\beta\alpha}$. The symmetric and antisymmetric parts of $\mathcal{Q}^{\alpha\beta}$ concerning the spatial indices define, respectively, the non-Abelian quantum metric and Berry curvature, $\mathcal{Q}^{\alpha\beta}=G^{\alpha\beta}+\frac{i}{2}\epsilon^{\alpha\beta}\Omega$, where $\epsilon^{\alpha\beta}$ is the antisymmetric tensor. The real and imaginary parts of Eq.~\eqref{fnlr2} are, 
\begin{equation}\label{sumrule1}
\int_0^{\infty} d\omega \frac{\text{Re}[\sigma_{\alpha\alpha}(\omega)]}{\omega}
=\frac{\pi}{2}\frac{e^2}{h}K,\quad\alpha=x \ \text{or}\ y,
\end{equation}
\begin{equation}\label{sumrule2}
\int_0^{\infty} d\omega \frac{\text{Im}[\sigma_{\alpha\beta}(\omega)]}{\omega}
=\frac{\pi}{2}\frac{e^2}{h}\epsilon^{\alpha\beta}C,\quad\alpha\neq\beta,
\end{equation}
where $K=\frac{1}{2\pi}\sum_{\alpha}\int d^2\bm{k}\text{Tr}~G^{\alpha\alpha}$ is identified as the quantum weight \cite{Onishi2024prx,OnishiQuantumweight2024} and $C=\frac{1}{2\pi}\int d^2\bm{k}\text{Tr}~\Omega$ is the Chern number.  Here the approximation of $|G\rangle$ being a Slater determinant is used, but no such approximation is required for the excited states $|\chi\rangle$. Since $|\chi\rangle$ can be exciton states induced by electron-hole attraction, Eqs.~(\ref{sumrule1}) and (\ref{sumrule2}) establish generalized sum rules for excitonic optical response in terms of quantum geometric quantities of ground state. 

The optical conductivity in response to circularly polarized light is $\sigma_{\pm}=\sigma_{xx} \pm i \sigma_{xy}$, of which the absorptive part is,
\begin{equation}\label{absor}
\text{Re} \sigma_{\pm}(\omega)=\text{Re}\sigma_{xx}(\omega)\mp\text{Im}\sigma_{xy}(\omega).
\end{equation}
The inequality $\text{Re} \sigma_{\pm}(\omega) \ge 0$ implies that $K \ge |C|$, which can be proved since the $2\times 2$ matrix $\Lambda$  with $\Lambda_{\alpha\beta}=\text{Tr}\mathcal{Q}^{\alpha\beta}$ is semipositive definite. When the bound is saturated ($K=|C|$),  either $\text{Re} \sigma_+ (\omega)$ or $\text{Re} \sigma_- (\omega)$ is strictly zero (depending on the sign of $C$), which leads to perfect MCD.

\begin{figure}[t]
\centering
\includegraphics[width=0.5\textwidth,trim=0 0 0 0,clip]{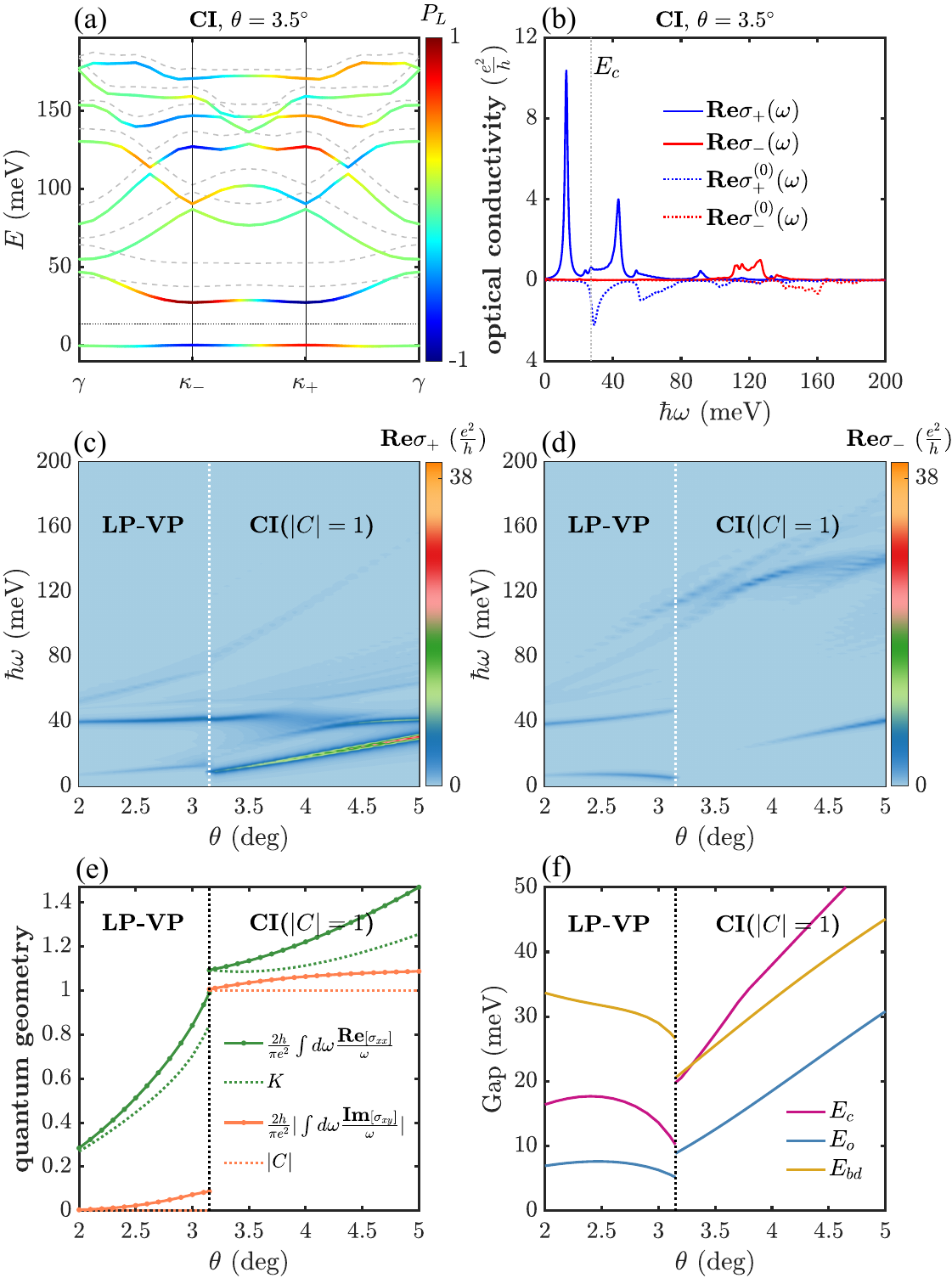}
\caption{(a) HF band structure of CI state at $\theta=3.5^\circ$ in the hole basis. The solid (long dashed) lines plot bands in $\tau=+ (-)$ valley. The dotted line marks the middle of the gap. The color represents the layer polarization $P_L$ with $+1$ $(-1)$ indicating the bottom (top) layer. (b) $\text{Re}\sigma_{\pm} (\omega)$ for the CI state in (a) . $\text{Re}\sigma^{(0)}_{\pm}(\omega)$ is calculated without electron-hole interaction. The vertical dotted line marks the minimum direct gap $E_c$. (c), (d) $\text{Re}\sigma_{\pm}(\omega)$ as a function of $\theta$ and $\hbar \omega$. White dotted lines separate the LP-VP and CI phases. (e) Quantum geometry obtained through optical sum rules (solid lines) and ground state (dashed lines). (f) The minimum direct gap $E_c$ (pink line), the  optical gap $E_o$ (blue line), and the  bound $E_{bd}$ (yellow line) as functions of $\theta$.
We take $\eta=1$ meV in the calculation of the absorption spectrum, following the experimental excitonic linewidth in $t$MoTe$_2$~\cite{Xiaodong2023b}.
}
\label{fig1}
\end{figure}

\textit{Excitonic optical response.---}We use the CI in $t$MoTe$_2$ at hole filling factor $\nu=1$, which has been experimentally realized \cite{Xiaodong2023a,Xiaodong2023b,Yihang2023_integer,Park2023b,Xu2023}, as an example to demonstrate the connection between excitonic optical response and quantum geometry.
In our calculation, we use a band-projected interacting model by retaining the top eight moir\'e valence bands in $t$MoTe$_2$, which ensures numerical convergence for physical quantities that we calculate (See Supplemental Material (SM)~\cite{suppl} for numerical details). The moir\'e bands are obtained by solving the continuum model \cite{Wu2019,MacDonaldRafiBistritzer2011} with the parameters from Ref.~\cite{WangChong2024}. Since we study holes doped into the system, we construct the projected Hamiltonian $\hat{\mathcal{H}}$ in the hole basis, which includes both the single-particle term $\hat{\mathcal{H}}_0$ and the Coulomb interaction term $\hat{\mathcal{H}}_\text{int}$,
\begin{equation}
\begin{aligned}
\hat{\mathcal{H}}_0&=\sum_{\boldsymbol{k}, \tau, n} \mathcal{E}_{\bm k}^{n\tau} b_{\boldsymbol{k} n \tau}^{\dagger} b_{\boldsymbol{k} n \tau},\\
\hat{\mathcal{H}}_\textrm{int}&=\frac{1}{2} \sum V_{\bm k_1 \bm k_2 \bm k_3 \bm k_4}^{n_1 n_2 n_3 n_4}\left(\tau, \tau'\right)
b_{\bm k_1 n_1 \tau}^{\dagger} b_{\bm k_2 n_2 \tau'}^{\dagger} b_{\bm k_3 n_3 \tau'} b_{\bm k_4 n_4 \tau}.
\end{aligned}
\label{H0Hint}
\end{equation}
Here $b_{\bm k n \tau}^{\dagger}$ ($b_{\bm k n \tau}$) is the hole creation (annihilation) operator for the $n$th moir\'e valence band at momentum $\bm k$ and valley $\tau$.  Here $\tau=+(-)$ is equivalent to spin up (down) \cite{Wu2019}.
$\mathcal{E}_{\bm k}^{n\tau}$ accounts for the single-particle band energy and  $V_{\bm k_1 \bm k_2 \bm k_3 \bm k_4}^{n_1 n_2 n_3 n_4}\left(\tau, \tau'\right)$ is the Coulomb potential $V_{\bm q}=2\pi e^2 \tanh{(|\bm q|d)}/(\epsilon |\bm q|)$ projected onto the moir\'e bands, where $d$ is the gate-to-sample distance and $\epsilon$ is the dielectric constant. In our calculation, we set $d=20$ nm, $\epsilon=20$ (See SM ~\cite{suppl} for more details).

We perform mean-field studies of $\hat{\mathcal{H}}$ using self-consistent Hartree-Fock (HF) approximation at $\nu=1$. We find two types of ground states separated by a first-order phase transition and tuned by the twist angle $\theta$ of $t$MoTe$_2$: (1) the CI phase with valley polarization but no layer polarization for $3.15^\circ <\theta < 5.0^\circ$ and (2) the phase with both layer polarization and valley polarization  (LP-VP phase) for $2.0^\circ <\theta < 3.15^\circ$. This phase diagram is consistent with previous studies \cite{InteractionDriven2023,Electricallytuned2024,MaximallyYangZhang2024}

\begin{figure}[t]
\centering
\includegraphics[width=0.5\textwidth,trim=0 0 0 0,clip]{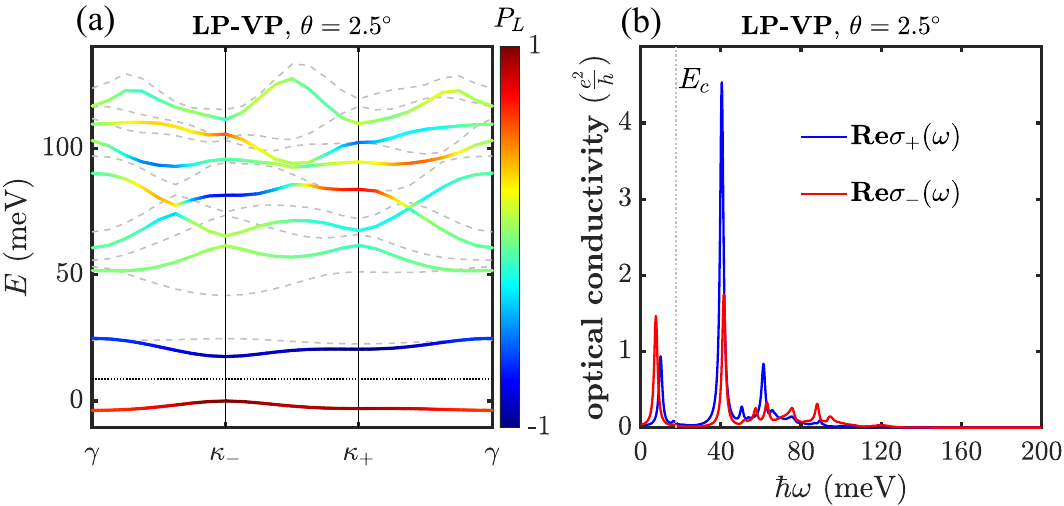}
\caption{(a) HF band structure of LP-VP state at $\theta=2.5^\circ$. (b) $\text{Re}\sigma_{\pm} (\omega)$ for the LP-VP state in (a).}
\label{fig2}
\end{figure}

The mean-field Hamiltonian for both  the CI and LP-VP phases can be formally  written as
\begin{equation}
\hat{\mathcal{H}}_{\text{MF}}=   \sum_{\boldsymbol{k}, \tau, \lambda} E_{\bm k}^{\lambda\tau} f_{\boldsymbol{k} \lambda \tau}^{\dagger} f_{\boldsymbol{k} \lambda \tau},
\end{equation}
where $E_{\bm k}^{\lambda\tau}$ includes the band energy and HF self energy. Here  $f_{\boldsymbol{k} \lambda \tau}^{\dagger}$ and $b_{\boldsymbol{k} n \tau}^{\dagger}$ operators are related by unitary transformations,  $f_{\boldsymbol{k} \lambda \tau}^\dagger=\sum_{n}U_{n \lambda }^{{\bm k} \tau}b_{\bm k n \tau}^{\dagger}$.
In Fig.~\ref{fig1}(a) [Fig.~\ref{fig2}(a)], we show the mean-field band structure for the  CI (LP-VP) phase, where the occupied band in the hole basis is layer hybridized (layer polarized) and carries a Chern number $C$ of $|C|=1(0)$. For definiteness, the occupied band is assumed to be polarized to the $\tau=+$ valley.

We now study intravalley excited states with zero center-of-mass momentum, which can be optically probed. In the presence of electron-hole interactions, the excited states can be parametrized as \cite{Excitonband2015}
\begin{equation}\label{exct}
|\chi\rangle=\sum_{\bm k,\lambda \ge 2}z_{\bm k,\lambda} (\chi) f_{\bm k\lambda+}^{\dagger} f_{\bm k1+}|G\rangle,
\end{equation}
where $|G\rangle$ is the Slate-determinant ground state obtained in the HF approximation, and  $|\chi\rangle$ is normalized with $\sum_{\bm k,\lambda \ge 2}|z_{\bm k,\lambda} (\chi)|^2=1$. Variation of the energy $\langle \chi | \hat{\mathcal{H}} |\chi \rangle$ with respect to the parameter $z_{\bm k,\lambda}(\chi)$ leads to the Bethe-Salpeter (BS) equation,
\begin{equation}\label{BS-equation1}
\begin{aligned}
&\mathcal{E}_{\chi}z_{\bm k,\lambda}(\chi)=\sum_{\bm p \lambda'}\mathcal{H}^{\lambda\lambda'}_{\bm k \bm p}z_{\bm p,\lambda'}(\chi), \\
&\mathcal{H}^{\lambda\lambda'}_{\bm k \bm p}=(E_{\bm k}^{\lambda +}- E_{\bm k}^{1 +})\delta_{\bm k \bm p}\delta_{\lambda\lambda'}
-(\tilde{V}^{1\lambda\lambda'1}_{\bm p\bm k\bm p\bm k}-\tilde{V}_{\bm k\bm p\bm p\bm k}^{\lambda 1 \lambda' 1}),
\end{aligned}
\end{equation}
where $\mathcal{H}^{\lambda\lambda'}_{\bm k \bm p}$ includes the quasiparticle energy cost of particle-hole transition as well as direct and exchange two-particle matrix element. Here $\tilde{V}$ denotes Coulomb matrix element in the basis of $f_{\boldsymbol{k} \lambda +}^{\dagger}$ operators.

We solve the BS equation to obtain $|\chi\rangle$ and $\mathcal{E}_\chi$, which are used to calculate the optical conductivity in Eq.~\eqref{kubo}. An optically active state, as an eigenstate of the $\hat{C}_{3z}$ symmetry, responds exclusively to either left or right circularly polarized light and therefore, is chiral. Because time-reversal symmetry is spontaneously broken by the valley polarization, excited states with opposite chiralities are energetically nondegenerate, but both types can be present in the spectrum. 
The calculated absorption spectrum $\text{Re} \sigma_{\pm}(\omega)$ for the CI phase at $\theta=3.5^{\circ}$, shown in Fig.~\ref{fig1}(b), has two noticeable features: (1) the spectrum exhibits nearly perfect MCD with vanishing  $\text{Re} \sigma_{-}$ for $\hbar \omega$ up to $110$meV; (2) the spectrum is dominated by a large chiral excitonic peak in $\text{Re} \sigma_{+}$ at an energy $E_o$ below $E_c$, where $E_o$ represents the optical gap and $E_c$ is the minimum direct gap in the HF band structure. 
In comparison, we also calculate the spectrum $\text{Re} \sigma^{(0)}_{\pm}(\omega)$ without including the electron-hole interactions, where the response is smeared out over energies above $E_c$ [Fig.~\ref{fig1}(b)].   
The dependence of $\text{Re} \sigma_{\pm}(\omega)$ on $\theta$ is shown in Figs.~\ref{fig1}(c) and \ref{fig1}(d), where the above two features are generic for the CI phase.

The CI phase with a nonzero Chern number should always have a finite MCD because of the sum rule for the Hall conductivity in 
Eq.~\eqref{sumrule2}. On the other hand, the nearly perfect MCD observed above is a special property of $t$MoTe$_2$, where $K$ and $|C|$  in the QAHE phase almost saturate the bound $K \ge |C|$ as shown in Fig.~\ref{fig1}(e). The deviation measured by $(K-|C|)/(K+|C|)$ is as low as 4\% at $\theta=3.5^{\circ}$, explaining the nearly vanishing $\text{Re} \sigma_{-}$ given that $C$ is $-1$ here. We note that $K=|C|$ is identified as the ideal trace condition which can favor fractionalized states \cite{WangJie2021}. 
By contrast, the LP-VP phase has $C=0$ and can have weak MCD according to Eq.~\eqref{sumrule2}. This is indeed the case as shown by the absorption spectrum in Fig.~\ref{fig2}(b) for the LP-VP phase, where $\text{Re} \sigma_{-}$ and $\text{Re} \sigma_{+}$ do not show a strong contrast. 
We note that the transition between the CI and LP-VP phases is first-order type without gap closing [Fig.~\ref{fig1}(e)], leading to the jump of the quantum geometric quantities across the transition  [Fig.~\ref{fig1}(f)].

\begin{figure}[t]
\centering
\includegraphics[width=0.5\textwidth,trim=0 0 0 0,clip]{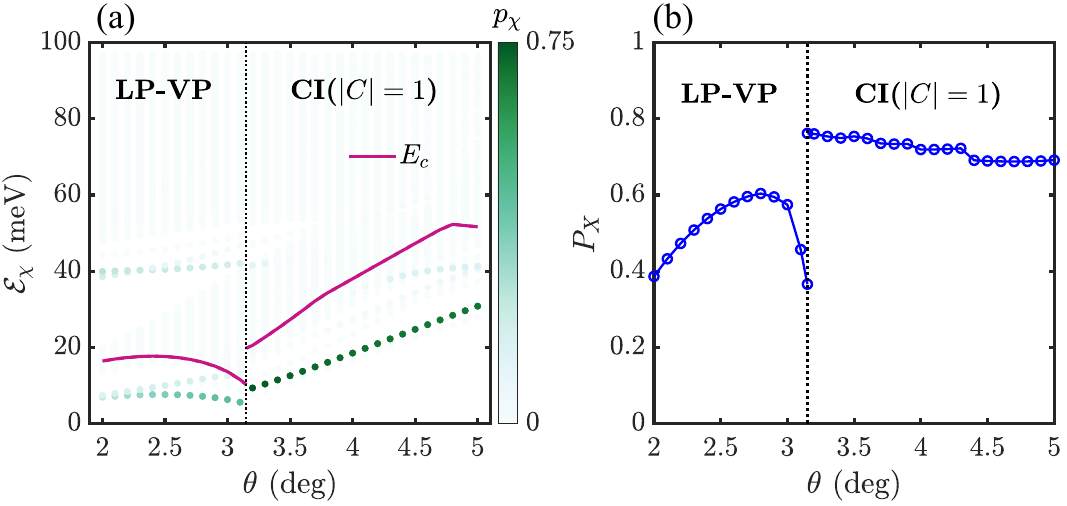}
\caption{(a) The color encodes $p_{\chi}$ for each excited state. The purple line marks  $E_c$. (b) $P_{X}$ as a function of $\theta$.}
\label{fig3}
\end{figure}

We compare values of $W_{\alpha \beta}^{(1)}$ calculated using, respectively, the excited states based on Eq.~\eqref{fnlr1} and the quantum geometry of ground state based on Eq.~\eqref{fnlr2}. The results obtained from the two approaches are not identical but have a semi-quantitative agreement, as plotted in Fig.~\ref{fig1}(e). The discrepancy comes from the fact that $|G\rangle$ and $|\chi\rangle$ used in our calculation are only approximate but not exact eigenstates of the interacting Hamiltonian $\hat{\mathcal{H}}$, while the sum rules are derived in the eigenbasis of  $\hat{\mathcal{H}}$. 

\begin{figure}[t]
\centering
\includegraphics[width=0.5\textwidth,trim=0 0 0 0,clip]{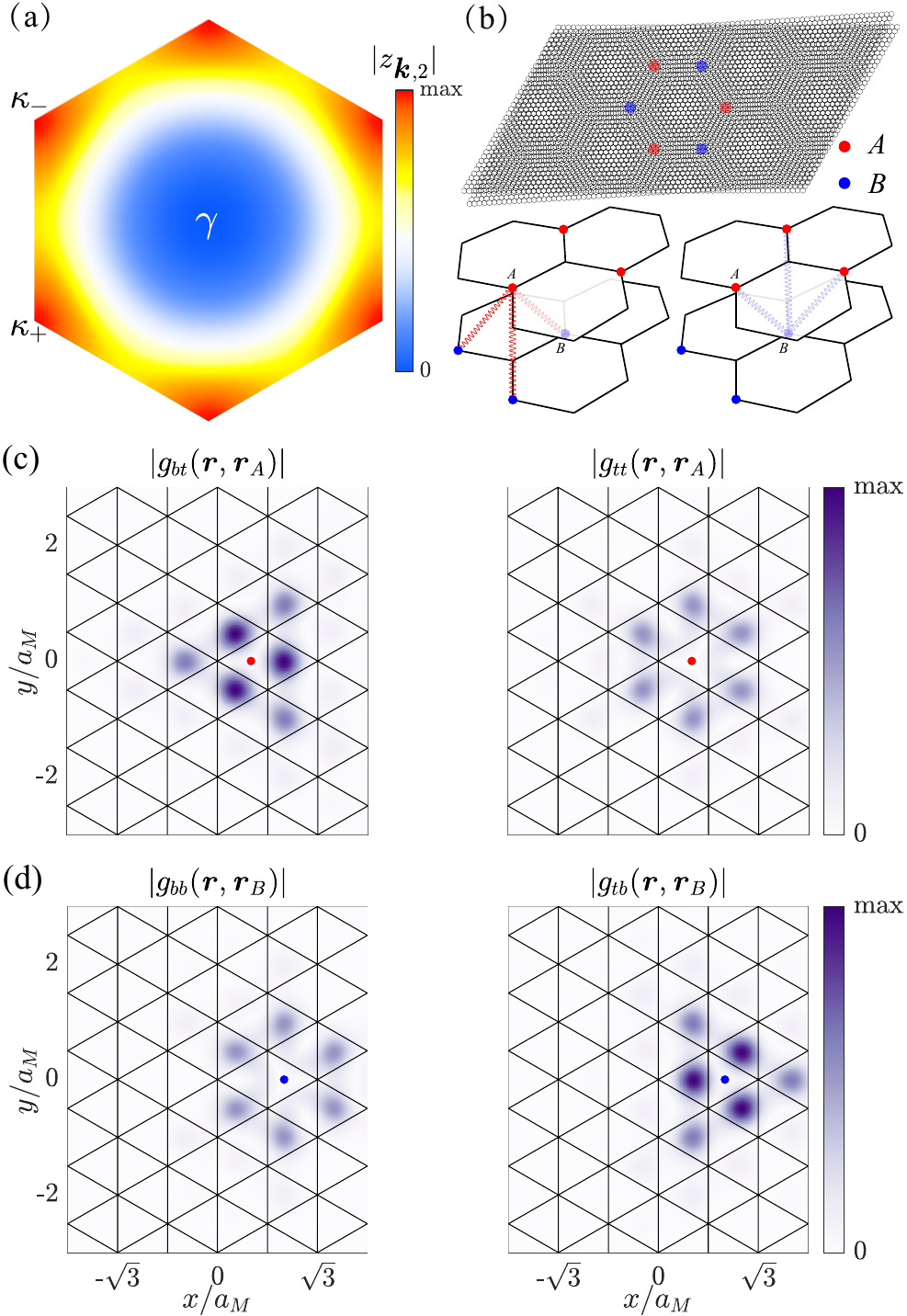}
\caption{Wavefunction of the dominating exciton in the CI phase at $\theta=3.5^\circ$. (a) $|z_{\bm k,2}|$ in the moir\'e Brillouin zone. (b) Top panel: Moir\'e superlattices of $t$MoTe$_2$ with $A$ and $B$ sites. Bottom panel: schematic illustration of interlayer excitations. (c), (d) $|g_{ll'}\left(\bm r, \bm r^{\prime}\right)|$ plotted in $\bm r$ space for different $l$ with fixed $\bm r'$ and $l'$. $a_M$ is the moir\'e period.}
\label{fig4}
\end{figure}

The exciton states can contribute significantly to $W_{\alpha \beta}^{(1)}$ because they have low energies and large optical matrix elements. To quantify the contribution of each excited state to $W_{xx}^{(1)}$, we define $p_{\chi}=\frac{w^{(1)}_{xx}(\chi)}{\sum_{\chi^\prime}w^{(1)}_{xx}(\chi^\prime)}$. As shown in Fig.~\ref{fig3}(a), the exciton states can have a sizable $p_{\chi}$. For example, the single dominating chiral exciton at $\theta=3.5^{\circ}$ in the CI phase has $p_{\chi}=0.73$. We further define $P_{X}=\sum_{\mathcal{E}_\chi<E_c}p_{\chi}$ to count the contribution of all exciton states below $E_c$, which can reach up to 0.8 (0.6) in the CI (LP-VP) phase [Fig.~\ref{fig3}(b)]. An interesting analogy is the optical response of quantum Hall states in Landau levels, which only occurs at the cyclotron resonance due to Kohn's theorem \cite{Kohn1961Cyclotron}.

An upper bound $E_{bd}$ can be put on the optical gap $E_o$ (the lowest energy of optically active excitons) by comparing Eq.~\eqref{sumrule1} with the $f$ sum rule given by (See SM~\cite{suppl} for derivation)
\begin{equation}\label{fsumrule}
\int_0^{\infty} d\omega \text{Re} \sigma_{xx}(\omega)=\frac{\pi n_c e^2}{2 m^*},
\end{equation}
where $m^*$ is the effective mass of the $t$MoTe$_2$ continumm model and $n_c$ is the carrier density. 
Using the inequality,
\begin{equation}\label{fsumrulebound}
\int_0^{\infty} d\omega \frac{\text{Re} \sigma_{xx}(\omega)}{\omega} \le \frac{\hbar}{E_o}\int_0^{\infty} d\omega \text{Re} \sigma_{xx}(\omega),
\end{equation}
we find the bound $E_o \le E_{bd}= \frac{2\pi{\hbar}^2n_c}{m^*K}$. Our numerical results indeed satisfy the bound $E_o \le E_{bd}$, as shown in Fig.~\ref{fig1}(f). However, $E_c$ obtained from the interaction-renormalized HF band structure, which would be the optical gap if electron-hole interactions were incorrectly neglected, is not bounded by $E_{bd}$ [Fig.~\ref{fig1}(f)]. Here $E_c$ provides an estimation of the mean-field charge gap. We note that an effective continuum Hamiltonian is used for $t$MoTe$_2$, and the frequency integral in the various sum rules should be understood as the integral over [$0$,$E_\text{max}/\hbar$], with $E_\text{max}$ being the largest energy scale that the effective Hamiltonian is still applicable \cite{Onishi2024prx,MaoDan2024}.

\textit{Exciton wave function.---}We examine the character of the dominating chiral exciton in the CI phase, which consists of particle-hole transitions in the $\tau=+$ valley mainly from the first band ($\lambda=1$) to the second band ($\lambda=2$).  The momentum-space wavefunction $|z_{\bm k,2}|$ of this exciton at $\theta=3.5^\circ$ is shown in Fig.~\ref{fig4}(a), which exhibits two peaks at the moir\'e Brillouin zone corners $\boldsymbol{\kappa}_+$ and $\boldsymbol{\kappa}_-$. Around $\boldsymbol{\kappa}_+$, the first and second bands are polarized, respectively, to the bottom ($b$) and top ($t$) layers of $t$MoTe$_2$, as illustrated in Fig.~\ref{fig1}(a). The situation is opposite around  $\boldsymbol{\kappa}_-$, where the first (second) band is polarized to the $t$ ($b$) layer. Therefore, the exciton comprises interlayer particle-hole transitions but has zero out-of-plane dipole moment due to the superposition of excitations at $\boldsymbol{\kappa}_+$ and $\boldsymbol{\kappa}_-$. 

The real-space exciton wavefunction $g_{ll'}(\bm r,\bm r')$ also exhibits interesting patterns, where $l$ ($l'$) is the layer index and $\bm r$ ($\bm r'$) is the in-plane position of the $p^*$ ($h^*$) particle. Here $p^*$ ($h^*$)  labels the particle (hole) in the hole basis that we employ.  Low-energy carriers in the $t$ and $b$ layer of $t$MoTe$_2$ are, respectively, confined to $A$ and $B$ sites in the moir\'e superlattice, which form a buckled honeycomb lattice~\cite{Wu2019,Devakul2021} [Fig.~\ref{fig4}(b)]. When $h^*$ is fixed in the $t$ layer at an $A$ site with position $\bm{r}_A$, the wavefunction $g_{l t}(\bm r,\bm{r}_A)$ is peaked at three nearest-neighbor $B$ sites in the $b$ layer, as plotted in Fig.~\ref{fig4}(c). Figure ~\ref{fig4}(d) shows a similar pattern for $g_{l b}(\bm r,\bm{r}_B)$. Therefore, the exciton has a charge transfer nature \cite{NaikIntralayer2022} in real space with the electron and hole separated by a length scale set by the moir\'e period.

\textit{Discussion.---}In summary, we present a theory that connects the quantum geometry of the ground state with the excitonic optical response through the optical sum rules.  We illustrate the implication of the quantum geometry on the excitonic optical response using the CI states in $t$MoTe$_2$ as an example. We further demonstrate that exciton states can dominate the optical sum rule, which is a quantum-mechanical phenomenon of coherent superposition in the exciton wavefunction, as explained in the SM \cite{suppl}.
This dominant contribution makes it practical to probe the quantum weight $K$ and Chern number $C$ using only the low-energy optical spectrum based on Eqs.~(\ref{sumrule1}) and (\ref{sumrule2}). We note that the Chern number $C$ can be independently measured by the Hall conductivity through current transport. Therefore, the quantum weight $K$ can be determined through the ratio $K/C$, which only requires relative values of $\text{Re} \sigma_{\pm} (\omega)$.   As we show in $t$MoTe$_2$, the relevant optical response of CI states is typically in the terahertz frequency range, which can be experimentally probed using available techniques \cite{LongJuTunable2017,Kato2023,RussellInfrared2023,LiangEvidence2024,BacSeulKiProbing2024,kumar2024terahertz,Ultrastrong2024}. 
Light-matter coupling in the terahertz regime is currently under active experimental study in van der Waals heterostructures \cite{kumar2024terahertz,Ultrastrong2024}. Extending beyond CIs, we anticipate that the quantum geometry can be explored through excitonic optical response for other topological states, for example, fractional Chern insulators.

\textit{Acknowledgments.---}We thank Lingjie Du and Yanhao Tang for their valuable discussions. This work is supported by National Key Research and Development Program of China (Grants No. 2021YFA1401300 and No. 2022YFA1402401), National Natural Science Foundation of China (Grant No. 12274333). W.-X. Q. is also supported by the China Postdoctoral Science Foundation (Grants No. 2024T170675 and No. 2023M742716). The numerical calculations in this paper have been done on the supercomputing system in the Supercomputing Center of Wuhan University.

\bibliography{reference}

\newpage
\onecolumngrid
\newtheorem{sectiontheorem}{Theorem}[section]
\newtheorem{corollary}{Corollary}[sectiontheorem]
\newtheorem{example}{Example}[sectiontheorem]
\renewcommand{\theequation}{S\arabic{equation}}
\renewcommand{\thefigure}{S\arabic{figure}}
\renewcommand{\thetable}{S\arabic{table}}
\setcounter{equation}{0}
\setcounter{figure}{0}
\setcounter{table}{0}
\setcounter{sectiontheorem}{0}

\begin{center}
\begin{large}
\textbf{Supplemental Material for ``Quantum Geometry Probed by Chiral Excitonic Optical Response of Chern Insulators"}
\end{large}
\end{center}

\section{Detailed derivation of the generalized optical sum rule}
The Kubo formula of Eq.~(1) in the main text is given by,
\begin{equation}\label{Kubo}
\begin{split}
\sigma_{\alpha \beta}(\omega)=&\sum_\chi F_{+}(\chi) V_{G \chi}^\alpha V_{\chi G}^\beta+F_{-}(\chi) V_{\chi G}^\alpha V_{G \chi}^\beta,\\
F_{\pm}(\chi)=&i\frac{e^2}{\hbar} \frac{1}{\mathcal{A}}\frac{1}{\mathcal{E}_\chi}\frac{1}{\hbar \omega \mp \mathcal{E} \chi+i \eta}.
\end{split}
\end{equation}
In the general case, the absorptive part of optical conductivity can be defined as \cite{Onishi2024prx},
\begin{equation}\label{sgmabs}
\begin{split}
\sigma^{\text{abs}}_{\alpha\beta}(\omega)=\text{Re}\sigma_{{\alpha\beta}}^L(\omega)+i\ \text{Im}\sigma_{{\alpha\beta}}^H(\omega),\quad \sigma_{{\alpha\beta}}^L(\omega)=\frac{\sigma_{{\alpha\beta}}(\omega)+\sigma_{{\beta\alpha}}(\omega)}{2},\quad \sigma_{{\alpha\beta}}^H(\omega)=\frac{\sigma_{{\alpha\beta}}(\omega)-\sigma_{{\beta\alpha}}(\omega)}{2}.
\end{split}
\end{equation}
From  Eq~\eqref{Kubo}, we obtain
\begin{equation}\label{sgmlh}
\begin{split}
\sigma_{{\alpha\beta}}^L(\omega)=\sum_{\chi}[F_{+}(\chi)+F_{-}(\chi)]\text{Re}[V_{G \chi}^\alpha V_{\chi G}^\beta],\quad
\sigma_{{\alpha\beta}}^H(\omega)=\sum_{\chi}i[F_{+}(\chi)-F_{-}(\chi)]\text{Im}[V_{G \chi}^\alpha V_{\chi G}^\beta],
\end{split}
\end{equation}
and
\begin{equation}\label{sgmabs2}
\begin{split}
\sigma^{\text{abs}}_{\alpha\beta}(\omega)=\sum_{\chi}V_{G \chi}^\alpha V_{\chi G}^\beta \text{Re}F_{+}(\chi)+(V_{G \chi}^\alpha V_{\chi G}^\beta)^* \text{Re}F_{-}(\chi).
\end{split}
\end{equation}
Using $\lim _{\eta \rightarrow 0^{+}} \frac{1}{x+i \eta}=P\left(\frac{1}{x}\right)-i \pi \delta(x)$,
\begin{equation}\label{f1f2}
\begin{split}
\text{Re}F_{\pm}(\chi)=\frac{e^2}{\hbar} \frac{\pi}{\mathcal{A}}\frac{1}{\mathcal{E}_\chi}\delta(\hbar \omega \mp \mathcal{E}_\chi).
\end{split}
\end{equation}
Therefore, the negative first  moment $W^{(1)}_{\alpha\beta}$ of $\sigma^{\text{abs}}_{\alpha\beta}(\omega)$ is calculated to be, 
\begin{equation}\label{fnlr1}
\begin{aligned}
W^{(1)}_{\alpha\beta}=&\int_0^{\infty} d \omega \frac{\sigma_{\alpha \beta}^{\mathrm{abs}}(\omega)}{\omega}=\frac{e^2}{\hbar} \frac{\pi}{\mathcal{A}} \sum_\chi w^{(1)}_{\alpha\beta}(\chi),
\end{aligned}
\end{equation}
where $w^{(1)}_{\alpha\beta}(\chi)=V_{G \chi}^\alpha V_{\chi G}^\beta/\mathcal{E}_\chi^2$.
By incorporating $\hat{\bm v}=i[\hat{\mathcal{H}},\bm \hat{\bm r}]$ into $V_{G \chi}^\alpha=\langle G|\hat{v}^{\alpha}|\chi \rangle$ and
$V_{\chi G}^\beta=\langle \chi|\hat{v}^{\beta}|G \rangle$, 
\begin{equation}\label{chisum}
\begin{aligned}
\sum_{\chi}w^{(1)}_{\alpha\beta}(\chi)=\sum_\chi \frac{\langle G|\hat{r}^\alpha\hat{\mathcal{H}}-\hat{\mathcal{H}}\hat{r}^\alpha|\chi \rangle\langle \chi|\hat{\mathcal{H}}\hat{r}^\beta-\hat{r}^\beta\hat{\mathcal{H}}|G \rangle}{\mathcal{E}_\chi^2}=\sum_\chi\langle G|\hat{r}^\alpha|\chi \rangle\langle \chi|\hat{r}^\beta|G \rangle=\langle G|\hat{r}^\alpha\hat{r}^\beta|G \rangle-\langle G|\hat{r}^\alpha|G \rangle\langle G|\hat{r}^\beta|G \rangle.
\end{aligned}
\end{equation}

We focus on the special case where $|G \rangle$ can be approximated by a Slater determinant composed of occupied Bloch states $|o \bm k \rangle$. Here $o$ is the band index and $\bm k$ labels momentum. The unoccupied  Bloch states are denoted as $|\bar{o} \bm k \rangle$. In the complete  basis of $\{|o \bm k \rangle, |\bar{o} \bm k \rangle\}$, the position operator $\bm \hat{\bm r}$ in the second quantized form is expressed as~\cite{Karplus1954,Ahn2022},
\begin{equation}\label{ropt}
\begin{aligned}
\hat{\bm r}(\bm k)=\sum_{\bm k} \sum_{mn}
\bm r_{mn}(\bm k) f_{\bm k,m}^{\dagger} f_{\bm k,n},\quad 
\bm r_{mn}(\bm k)=\delta_{mn} i \partial_{\bm k}-\bm A_{mn}(\bm k),\quad \bm A_{mn}(\bm k)=-\left\langle u_{m \bm k}\left|i \partial_{\bm k}\right|u_{n \bm k}\right\rangle,
\end{aligned}
\end{equation}
where $|u_{n \bm k}\rangle$ is the periodic part of the Bloch state $|n \bm k\rangle $ and $\bm A_{mn}(\bm k)$ is the Berry connection. Then
\begin{equation}\label{rmatrix}
\begin{aligned}
\langle G|\hat{r}^\alpha\hat{r}^\beta|G \rangle=&\sum_{\bm k_1} \sum_{m_1 n_1}\sum_{\bm k_2} \sum_{m_2 n_2} r_{m_1 n_1}^{\alpha}(\bm k_1) r_{m_2 n_2}^{\beta}(\bm k_2) \langle G|f_{\bm k_1,m_1}^{\dagger} f_{\bm k_1,n_1}f_{\bm k_2,m_2}^{\dagger} f_{\bm k_2,n_2}|G \rangle \\
=&\sum_{\bm k_1 \bm k_2} \sum_{o_1o_2} r_{o_1 o_1}^{\alpha}(\bm k_1) r_{o_2 o_2}^{\beta}(\bm k_2)
+\sum_{\bm k}\sum_{o\bar{o}} r_{o \bar{o}}^{\alpha}(\bm k) r_{\bar{o} o}^{\beta}(\bm k),\\
\langle G|\hat{r}^\alpha|G \rangle\langle G|\hat{r}^\beta&|G \rangle=\sum_{\bm k_1 \bm k_2} \sum_{o_1o_2} r_{o_1 o_1}^{\alpha}(\bm k_1) r_{o_2 o_2}^{\beta}(\bm k_2).
\end{aligned}
\end{equation}

By combining Eqs.~\eqref{chisum}, \eqref{ropt} and \eqref{rmatrix}, Eq.~\eqref{fnlr1} can be further written as,
\begin{equation}
\begin{aligned}
W^{(1)}_{\alpha\beta}=\frac{e^2}{\hbar} \frac{\pi}{\mathcal{A}}\sum_{\bm k}\sum_{o\bar{o}} r_{o \bar{o}}^{\alpha}(\bm k) r_{\bar{o} o}^{\beta}(\bm k)
=\frac{e^2}{\hbar} \frac{\pi}{\mathcal{A}}\sum_{\bm k}\sum_{o\bar{o}} A_{o \bar{o}}^{\alpha}(\bm k) A_{\bar{o} o}^{\beta}(\bm k)
=\frac{e^2}{\hbar} \frac{\pi}{\mathcal{A}}\sum_{\bm k}\text{Tr}\mathcal{Q}^{\alpha\beta},
\end{aligned}
\end{equation}
where $\mathcal{Q}^{\alpha\beta}_{oo'}=\sum_{\bar{o}} A_{o \bar{o}}^\alpha A_{\bar{o} o'}^\beta$ is the quantum geometric tensor of occupied bands \cite{MaYuQuan2010}.

To study the response to circularly polarized light, we define the velocity operator $\hat{v}^{\pm}=\frac{1}{\sqrt{2}}\left(\hat{v}^x \pm i \hat{v}^y\right)$ and the optical matrix element $V_{\chi G}^{\pm}= \langle 
 \chi |\hat{v}^{\pm}|G\rangle$. The corresponding optical conductivity is given by,
\begin{equation}
\begin{aligned}
\sigma_{ \pm}(\omega)& =i \frac{e^2}{\hbar} \frac{1}{\mathcal{A}} \sum_\chi \frac{1}{\mathcal{E}_\chi}\left[\frac{V_{G \chi}^{\mp} V_{\chi G}^{ \pm}}{\hbar \omega-\mathcal{E}_\chi+i \eta}+\frac{V_{\chi G}^{ \mp} V_{G \chi}^{\pm}}{\hbar \omega+\mathcal{E}_\chi+i \eta}\right] \\
& =i \frac{e^2}{2\hbar} \frac{1}{\mathcal{A}} \sum_\chi \frac{1}{\mathcal{E}_\chi}\left[\frac{\left(V_{G \chi}^x V_{\chi G}^x+V_{G \chi}^y V_{\chi G}^y \pm i V_{G \chi}^x V_{\chi G}^y \mp i
V_{G \chi}^y V_{\chi G}^x \right)}{\hbar \omega-\mathcal{E}_\chi+i \eta}+\frac{\left(V_{\chi G}^x V_{G \chi }^x+V_{\chi G}^y V_{G \chi }^y \pm i V_{\chi G}^x V_{G \chi }^y \mp i
V_{\chi G}^y V_{G \chi}^x \right)}{\hbar \omega+\mathcal{E}_\chi+i \eta}\right] \\
& =\frac{1}{2}[\sigma_{x x}(\omega)+\sigma_{y y}(\omega) \pm i \sigma_{x y}(\omega) \mp i \sigma_{y x}(\omega)].
\end{aligned}
\end{equation}
We focus on two-dimensional systems with an out-of-plane threefold rotational symmetry $\hat{C}_{3z}$, where $\sigma_{xx}=\sigma_{yy}$ and $\sigma_{xy}=-\sigma_{yx}$,
which leads to Eq.~(7) in the main text.

\section{Model Hamiltonian and ground state calculations}
\subsection{Moir\'e Hamiltonian}
Moir\'e superlattices of $t$MoTe$_2$ respect a threefold rotation $\hat{C}_{3z}$ around the out-of-plane $\hat{z}$ axis and a twofold rotation $\hat{C}_{2y}$ around the in-plane $\hat{y}$ axis that exchanges the bottom ($b$) and top ($t$) layers.
The single-particle moir\'e Hamiltonian of $t$MoTe$_2$ has been constructed in Ref.~\cite{Wu2019} for valence band states in $\pm K$ valley as,
\begin{equation}\label{ctnh0}
\begin{aligned}
&\hat{\mathcal{H}}^{\tau}_0  =\left(\begin{array}{cc}
-\frac{\hbar^2\left(\hat{\boldsymbol{k}}-\tau\boldsymbol{\bm \kappa}_{+}\right)^2}{2 m^*}+\Delta_{+}(\boldsymbol{r}) & \Delta_{\mathrm{T,\tau}}(\boldsymbol{r}) \\
\Delta_{\mathrm{T,\tau}}^{\dagger}(\boldsymbol{r}) & -\frac{\hbar^2\left(\hat{\boldsymbol{k}}-\tau\bm\kappa_{-}\right)^2}{2 m^*}+\Delta_{-}(\boldsymbol{r})
\end{array}\right), \\
&\Delta_{\pm}(\boldsymbol{r}) = 2 V \sum_{j=1,3,5} \cos \left(\boldsymbol{g}_j \cdot \boldsymbol{r} \pm \psi\right), \\
&\Delta_{\mathrm{T,\tau}}(\boldsymbol{r})=w \left(1+e^{-i \tau\bm g_2 \cdot \boldsymbol{r}}+e^{-i \tau \bm g_3 \cdot \boldsymbol{r}}\right),
\end{aligned}
\end{equation}
where the $2\times2$ Hamiltonia $\hat{\mathcal{H}}^{\tau}_0$ is expressed in the layer-pseudospin space. The index $\tau=\pm$ labels $\pm K$ valleys, which are also locked to spin $\uparrow$ and $\downarrow$, respectively.
$\Delta_\pm (\bm{r})$ is the layer-dependent potential with an amplitude $V$ and phase parameters $\pm \psi$, and $\Delta_{\mathrm{T,\tau}}(\boldsymbol{r})$ is the interlayer tunneling with a strength $w$. $\bm{r}$ and  $\hat{\bm{k}}=-i\partial_{\bm r}$ are respectively, the position and momentum operators. $m^*$ is the effective mass. $\bm{\kappa}_{\pm}=\left[4\pi /(3 a_M)\right](-\sqrt{3}/2, \mp 1/2 )$ are located at corners of the moir\'e Brillouin zone, and $\bm{g}_j=\left[4\pi /(\sqrt{3} a_M)\right]\{\cos[(j-1)\pi/3], \sin[(j-1)\pi/3]\}$ for $j=1,...,6$ are the moir\'e reciprocal lattice vectors, where $a_M\approx a_0/\theta$ is the moir\'e period and $a_0$ is the monolayer lattice constant.

In this work, we focus on the Chern insulator (CI) phase for $\theta$ around $3.5^\circ$, where the Bloch states have a nearly ideal quantum geometry. For this range of $\theta$, appropriate parameters can be taken as, $a_0 = 3.52 $\AA, $m^* = 0.6 m_e$, $V = 20.8$ meV, $\psi = -107.7^\circ$, $w = -23.8$ meV, as used in Refs.~\cite{WangChong2024,LiuXiaoyu2024,FanFengRen2024}, where $m_e$ is the electron bare mass. Under these parameters, the first two bands have opposite Chern numbers in each valley and can be mapped to Kane-Mele model on a honeycomb lattice composed of $A$ and $B$ sites~\cite{Wu2019,Devakul2021}.

\subsection{Band-projected Interacting Hamiltonian}
The moir\'e Hamiltonian $\hat{\mathcal{H}}^{\tau}_0$ in Eq.~\eqref{ctnh0} can be solved within plane wave basis to obtain the energy $\varepsilon_{\bm k}^{n\tau}$ and wave function $\phi_{\bm k }^{n \tau}\left(r\right)$ for the $n$-th moir\'e band at $\bm k$ and $\tau$. The interacting model is then constructed under the moir\'e band basis. Because all valence band states are below the Fermi energy for the charge-neutral twisted homobilayer, it is more convenient to use the hole basis for hole doped system. We define the hole operator as $b_{\boldsymbol{k}n\tau}=c_{\boldsymbol{k}n\tau}^{\dagger}$, where $c_{\boldsymbol{k}n\tau}^{\dagger}$ is the creation operator for the Bloch state $\phi_{\bm k }^{n \tau}\left(r\right)$. The single-particle Hamiltonian $\hat{\mathcal{H}}_0$ in the hole basis can be written as
\begin{equation}
\label{H0}
\hat{\mathcal{H}}_0=\sum_{\boldsymbol{k}, \tau, n} \mathcal{E}_{\bm k}^{n\tau} b_{\boldsymbol{k} n \tau}^{\dagger} b_{\boldsymbol{k} n \tau},
\end{equation}
where $\mathcal{E}_{\bm k}^{n\tau}=-\varepsilon_{\bm k}^{n\tau}$ is the energy of Bloch state. In the hole representation, the Coulomb interaction projected onto the  moir\'e bands is expressed as 
\begin{equation}
\label{Hint}
\begin{split}
\hat{\mathcal{H}}_\textrm{int}=\frac{1}{2} \sum V_{\bm k_1 \bm k_2 \bm k_3 \bm k_4}^{n_1 n_2 n_3 n_4}\left(\tau, \tau'\right)
b_{\bm k_1 n_1 \tau}^{\dagger} b_{\bm k_2 n_2 \tau'}^{\dagger} b_{\bm k_3 n_3 \tau'} b_{\bm k_4 n_4 \tau},
\end{split}
\end{equation}
where the summation is over the momentum $\bm{k}_j$ (summed over the moir\'e Brillouin zone), the moir\'e band index $n_j$, and the valley index $\tau$.
The Coulomb matrix element is given by,
\begin{equation}
V_{\bm k_1 \bm k_2 \bm k_3 \bm k_4}^{n_1 n_2 n_3 n_4}\left(\tau, \tau'\right)=\frac{1}{\mathcal{A}} \sum_{\bm q} V_{\bm q} M_{\bm k_1 \bm k_4}^{n_1 n_4}\left(\tau, \bm q\right) M_{\bm k_2 \bm k_3}^{n_2 n_3}\left(\tau',-\bm q\right),
\end{equation}
where $\mathcal{A}$ is the system area.
The structure factor $M_{\bm k_1 \bm k_4}^{n_1 n_4}\left(\tau, \bm q\right)$ is 
\begin{equation}
M_{\bm k_1 \bm k_4}^{n_1 n_4}\left(\tau, \bm q\right)=\sum_{l} \int d\bm r e^{i \bm q \cdot\bm r} {[{\tilde{\phi}}_{\bm k_1 l}^{n_1 \tau}\left(\bm r\right)]}^{*} {\tilde{\phi}}_{\bm k_4 l}^{n_4 \tau}\left(\bm r\right),
\end{equation}
where $\tilde{\phi}_{\bm k }^{n \tau}\left(r\right)=[\phi_{\bm k }^{n \tau}\left(r\right)]^*$ due to the particle-hole transformation and $l$ is the layer index. We use the dual-gate screened Coulomb interaction with the momentum-dependent potential $V_{\bm q}=2\pi e^2 \tanh{(|\bm q|d)}/(\epsilon |\bm q|)$, where $d$ is the gate-to-sample distance and $\epsilon$ is the dielectric constant. In our calculation, we set $d=20$ nm, $\epsilon=20$.

The full Hamiltonian $\hat{\mathcal{H}}$ of the interacting system is,
\begin{equation}
\begin{split}
\hat{\mathcal{H}}=\hat{\mathcal{H}}_0+\hat{\mathcal{H}}_\textrm{int}. 
\end{split}
\label{H3}
\end{equation}
This Hamiltonian, written in the hole basis, takes into account the Coulomb interactions between holes. In this scheme, at charge neutrality point with zero hole doping, the band structure from the continuum moir\'e Hamiltonian subjects to no further interaction effects, as it is assumed to approximate the density-functional-theory band structure that already includes interaction effects. Double counting of the Coulomb interaction is avoided by working in the hole basis.

\subsection{Mean-field Calculation}
Based on Hartree-Fock approximation of $\hat{\mathcal{H}}_{\text{int}}$, we perform self-consistent calculations to study the ground states of $\hat{\mathcal{H}}$ for integer filling factors $\nu=1$ at zero temperature. Here $\nu=\frac{1}{N_k}\sum_{\bm k, n, \tau} b_{\bm k n \tau}^{\dagger} b_{\bm k n \tau}$ is the number of holes per moir\'e unit cell, and $N_k$ is the number of $\bm k$ points in the summation.
Starting from various symmetry-broken mean-field ansatzes, we self-consistently generate different mean-field solutions and compare their energy to determine the mean-field ground state. We calculate the phase diagram as a function of the twist angle $\theta$, with the gate-to-sample distance $d$ and the dielectric constant $\epsilon$ fixed. The parameter $d$ can be controlled by the thickness of the encapsulating hBN layer. The dielectric constant $\epsilon$ accounts for the environmental screening from hBN as well as internal screening from remote moir\'e bands. Here we take $\epsilon$ as a phenomenological parameter.

\section{Calculation of optical matrix element}
The velocity operator $\hat{\bm v}$ can be expressed as follows,
\begin{equation}\label{vecopt}
\begin{aligned}
&\hat{\bm v}=\sum_{\bm k,\tau} \sum_{m n} \bm J^{\tau}_{mn}(\bm k)
c_{\bm k m \tau}^{+} c_{\bm k n \tau}=
-\sum_{\bm k,\tau} \sum_{m n}[\bm J^{\tau}_{mn}(\bm k)]^*b_{\bm k m \tau}^{+} b_{\bm k n \tau}
=\sum_{\bm k,\tau} \sum_{\lambda \lambda^{\prime}} \bm v_{\tau,\lambda \lambda^{\prime}}(\bm k) f_{\bm k \lambda\tau}^{\dagger} f_{\bm k \lambda^{\prime}\tau},\\
&\bm J^{\tau}_{mn}(\bm k)=\langle\phi^{m\tau}_{\bm k} |
\frac{ \partial\hat{\mathcal{H}}^{\tau}_0 }{\partial{\hat{\bm k}}}| \phi^{n\tau}_{\bm k}\rangle, \\
&\bm v_{\tau,\lambda \lambda^{\prime}}(\bm k)=-\left[U_{m \lambda}^{\bm k\tau}\right]^*[\bm J^{\tau}_{mn}(\bm k)]^*
 U_{n \lambda^{\prime}}^{\bm k\tau},
\end{aligned}
\end{equation}
where $f_{\bm k\lambda\tau}^{\dagger}(f_{\bm k\lambda\tau})$ is the creation (annihilation) operator of Hartree-Fock quasiparticles of the $\lambda$-th band at $\bm k$ of valley $\tau$. Here $f_{\boldsymbol{k} \lambda \tau}^{\dagger}$ and $b_{\boldsymbol{k} n \tau}^{\dagger}$ operators are related by unitary transformations,  $f_{\boldsymbol{k} \lambda \tau}^\dagger=\sum_{n}U_{n \lambda }^{{\bm k} \tau} b_{\boldsymbol{k} n \tau}^{\dagger}$. According to Eq.~\eqref{ctnh0}, $\frac{ \partial\hat{\mathcal{H}}^{\tau}_0 }{\partial{\hat{\bm k}}}$ can be expressed as,
\begin{equation}\label{partialctnh0}
\begin{aligned}
\frac{ \partial\hat{\mathcal{H}}^{\tau}_0 }{\partial{\hat{\bm k}}}=\left(\begin{array}{cc}
-\frac{\hbar^2\left(\hat{\bm k}-\tau\boldsymbol{\bm \kappa}_{+}\right)}{m^*} & 0 \\
0 & -\frac{\hbar^2\left(\hat{\bm k}-\tau\boldsymbol{\bm \kappa}_{-}\right)}{m^*}
\end{array}\right).
\end{aligned}
\end{equation}

Using Eq.~\eqref{vecopt} and the exciton wave function of Eq.~(10) in the main text, we derive the optical matrix element as,
\begin{equation}\label{opmel}
\langle\chi|\hat{\bm v}| G\rangle=\sum_{\bm k^{\prime}, \lambda^{\prime \prime}\geq2} \sum_{\bm k, \lambda \lambda^{\prime}}\left[z_{\bm k^{\prime},\lambda^{\prime\prime}}( \chi)\right]^* \bm v_{+,\lambda \lambda^{\prime}}(\bm k)\left\langle G\left|f_{\bm k^{\prime}1+}^{\dagger} f_{\bm k^{\prime} \lambda^{\prime\prime}+} f_{\bm k\lambda+}^{\dagger} f_{\bm k\lambda^{\prime}+}\right| G\right\rangle=\sum_{\bm k, \lambda\geq2}\left[z_{\bm k,\lambda}(\chi)\right]^* \bm v_{+,\lambda 1}(\bm k).
\end{equation}
Equation ~\eqref{opmel} is used in the calculation of the optical conductivity.

\section{Derivation of $f$ sum rule}
Here, we also give the derivation of the $f$ sum rule presented in Eq.~(12) in the main text. According to Eq.~\eqref{Kubo}, we obtain
\begin{equation}\label{fsumrule}
W_{\alpha \alpha}^{(0)}=\int_0^{\infty} d\omega \text{Re} \sigma_{\alpha\alpha} (\omega)=\frac{e^2}{\hbar}\frac{\pi}{\mathcal{A}}\sum_{\chi}w^{(0)}_{\alpha\alpha}(\chi),
\end{equation}
where $w^{(0)}_{\alpha\alpha}(\chi)=V_{G \chi}^\alpha V_{\chi G}^\alpha/\mathcal{E}_\chi$. By incorporating $\hat{\bm v}=i[\hat{\mathcal{H}},\bm \hat{\bm r}]$ into $V_{G \chi}^\alpha=\langle G|\hat{v}^{\alpha}|\chi \rangle$ and
$V_{\chi G}^\alpha=\langle \chi|\hat{v}^{\alpha}|G \rangle$, 
\begin{equation}\label{chisum2}
\begin{aligned}
\sum_\chi w^{(0)}_{\alpha\alpha}(\chi)=
&\frac{i}{2}\sum_{\chi}\frac{\langle G|\hat{\mathcal{H}} \hat{r}^\alpha-\hat{r}^\alpha\hat{\mathcal{H}}|\chi\rangle\langle\chi|\hat{v}^{\alpha}|G\rangle}{\mathcal{E}_\chi}
+\frac{i}{2}\sum_{\chi}\frac{\langle G|\hat{v}^{\alpha}|\chi\rangle\langle\chi|\hat{\mathcal{H}} \hat{r}^\alpha-\hat{r}^\alpha\hat{\mathcal{H}}|G\rangle}{\mathcal{E}_\chi}\\
=&-\frac{i}{2}\sum_{\chi}\langle G|\hat{r}^\alpha|\chi\rangle\langle\chi|\hat{v}^{\alpha}|G\rangle
+\frac{i}{2}\sum_{\chi}\langle G|\hat{v}^{\alpha}|\chi\rangle\langle\chi|\hat{r}^{\alpha}|G\rangle\\
=&-\frac{i}{2}\langle G|[\hat{r}^\alpha,\hat{v}^{\alpha}]|G\rangle
=\frac{\hbar N_c}{2m^*},
\end{aligned}
\end{equation}
where $N_c$ is the total number of carriers. Then Eq.~\eqref{fsumrule} can be written as
\begin{equation}
W_{\alpha \alpha}^{(0)}=\int_0^{\infty} d\omega \text{Re} \sigma_{\alpha\alpha}=\frac{\pi n_c e^2}{2m^*}.
\end{equation}

The charge density $n_c$ at $\nu=1$ is $1/\mathcal{A}_0$, where $\mathcal{A}_0=\frac{\sqrt{3}}{2}a_M^2$ is the area of moir\'e unit cell.

\section{Real-space exciton wavefunction}
The exciton state in momentum-space is parametrized as,
\begin{equation}\label{exctsm1}
|\chi\rangle=\sum_{\bm k,\lambda \ge 2}z_{\bm k,\lambda} (\chi) f_{\bm k\lambda+}^{\dagger} f_{\bm k1+}|G\rangle.
\end{equation}
We expand $f_{\bm k\lambda+}^{\dagger}$ ($f_{\bm k1+}$)  by field operators at real space position $\bm r$ in layer $l$ as
\begin{equation}\label{fkrexp}
f_{\bm k\lambda+}^{\dagger}=\sum_{l}\int d\bm r\psi^{\lambda+}_{\bm kl}(\bm r) f_{l\bm r}^{\dagger},\quad
f_{\bm k1+}=\sum_{l}\int d\bm r[\psi^{1+}_{\bm kl}(\bm r)]^* f_{l\bm r}.
\end{equation}
Then Eq.~\eqref{exctsm1} can be written as
\begin{equation}\label{exctsm2}
|\chi\rangle=\sum_{l,l'}\int d\bm r \int d\bm r' g_{ll'}^{\chi}(\bm r,\bm r') f_{l\bm r}^{\dagger} f_{l'\bm r'}|G\rangle,
\end{equation}
where $g_{ll'}^{\chi}(\bm r,\bm r')=\sum_{\bm k,\lambda \ge 2}z_{\bm k,\lambda} (\chi) \psi^{\lambda+}_{\bm kl}(\bm r) [\psi^{1+}_{\bm kl'}(\bm r')]^*$ is the real space  exciton wavefunciton. Here $\psi^{\lambda+}_{\bm kl}(\bm r)$ ($\psi^{1+}_{\bm kl}(\bm r)$) is the quasiparticle Bloch wavefunction at valley $\tau=+$ obtained by the mean-field calculation.

\section{Numerical Convergence}
\begin{figure}[t]
\centering
\includegraphics[width=1.0\textwidth,trim=0 0 0 0,clip]{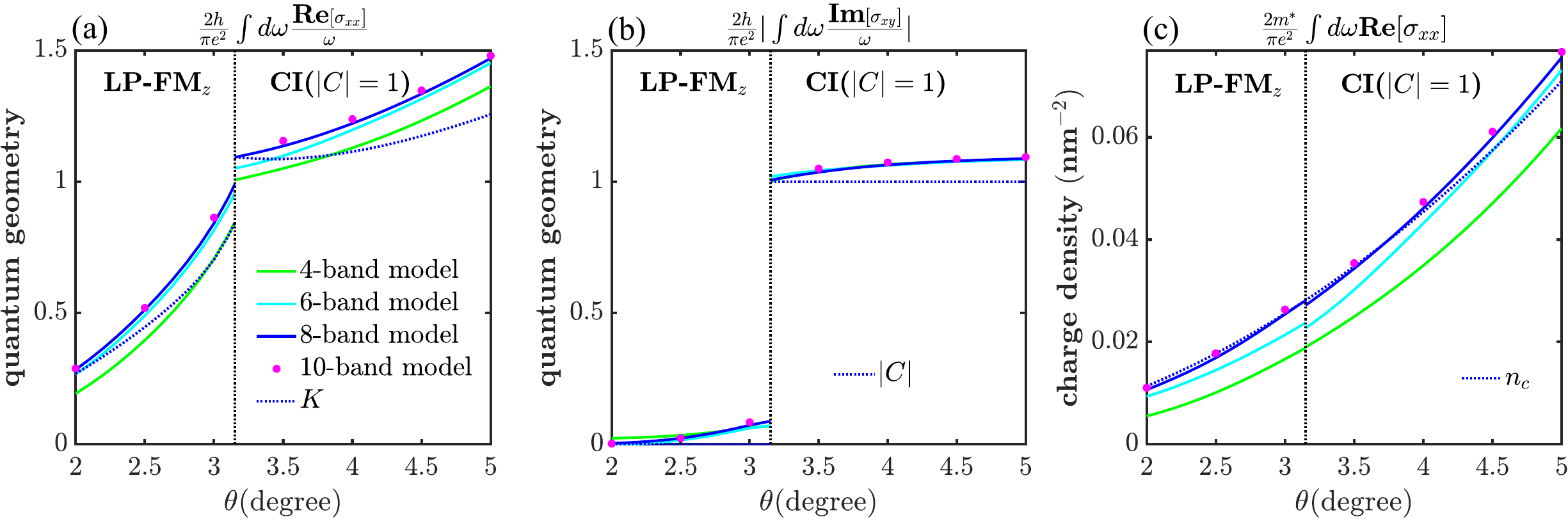}
\caption{Numerical results of optical sum rules calculated by including different numbers of moir\'e valence bands. (a), (b) The generalized optical weight $W^{(1)}_{\alpha\beta}$ with its real part in (a) and imaginary part in (b). (c) The optical spectral weight $W^{(0)}_{\alpha\beta}$. The results in (a), (b) and (c) are compared with quantum weight $K$ (calculated using 8 moir\'e bands), Chern number $|C|$, and charge density $n_c$, respectively. We use $24\times 24$ $k$-mesh in the calculation.}
\label{fig_fsr}
\end{figure}

In principle, the sum rules should include excited states over all energies. However, as shown in Fig.~3 (a) in our main text, contributions from excited states with high energies can be ignored. This implies that the band-projected interacting model of Eq.~\eqref{Hint} can be truncated by including only a finite number of moir\'e bands. In Fig.~\ref{fig_fsr}, we compare the numerical results of both $W^{(1)}_{\alpha\beta}$ and $W^{(0)}_{\alpha \alpha}$ calculated by including different numbers of bands. We find that numerical convergence is reached by including 8 moir\'e bands. 

\begin{figure}[t]
\centering
\includegraphics[width=1.0\textwidth,trim=0 0 0 0,clip]{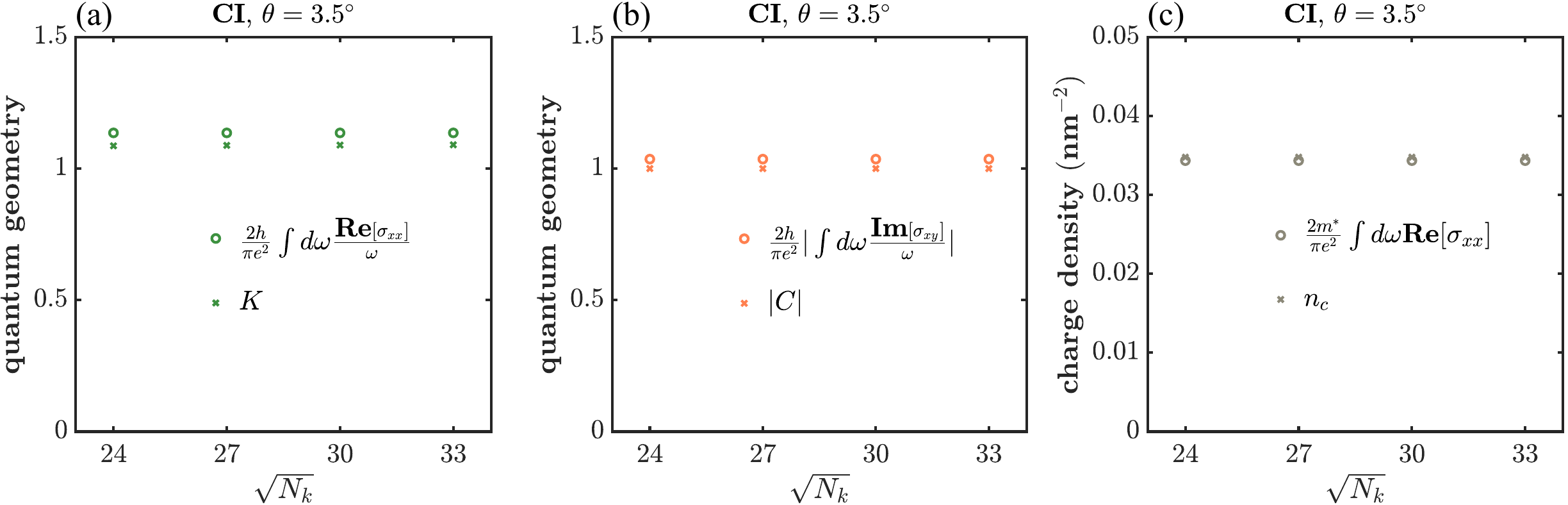}
\caption{Numerical results of optical sum rules for CI phase at $\theta=3.5^\circ$ calculated using different sizes of $k$-mesh. (a), (b) The generalized optical weight $W^{(1)}_{\alpha\beta}$ with its real part in (a) and imaginary part in (b). (c) The optical spectral weight $W^{(0)}_{\alpha \alpha}$. The results in (a), (b), and (c) are compared with quantum weight $K$, Chern number $|C|$, and charge density $n_c$, respectively. We keep $8$ moir\'e bands in the calculation.} 
\label{fig_vnk}
\end{figure}

We also compare the numerical results calculated using different sizes of $k$-mesh, as shown in Fig.~\ref{fig_vnk}. We find that $N_k=24\times 24$ is enough for numerical convergence.

\section{Physical Discussion}
\subsection{Sum rule within a projected subspace}
The sum rule in Eq.~(2) of main text is constructed by carrying out an integral to infinity, while the model study in $t$MoTe$_2$ is performed within a projected subspace by retaining only the “valence” states. With the use of the effective continuum Hamiltonian description, the frequency integral in the various sum rules should be understood as the integral over [$0$,$E_\text{max}/\hbar$], with $E_\text{max}$ being the largest energy scale that the effective Hamiltonian is still applicable \cite{Onishi2024prx,MaoDan2024}. For our model study of $t$MoTe$_2$, $E_\text{max}$ is about 200 meV, above which the optical absorption within the effective model nearly vanishes.

\begin{figure}[b]
\centering
\includegraphics[width=0.7\textwidth,trim=0 0 0 0,clip]{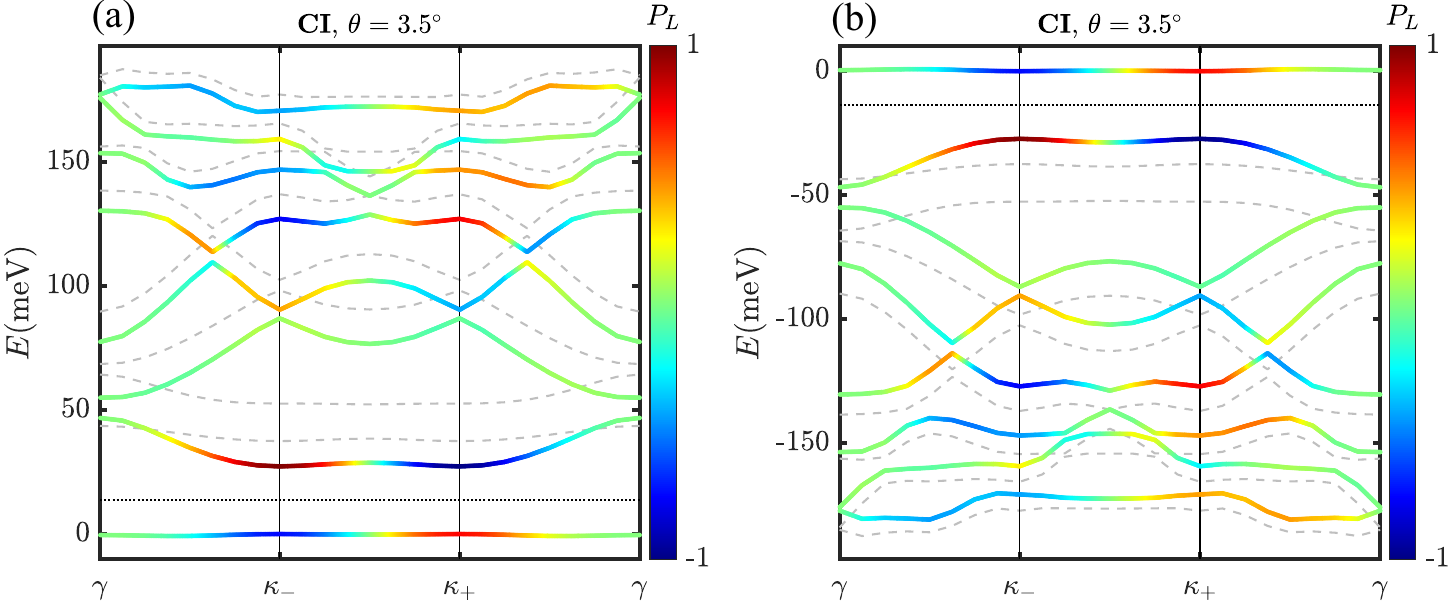}
\caption{The self-consistent Hartree-Fock band structure at hole filling $\nu=1$ in (a) the hole basis and (b) the physical electron basis, which are related by a particle-hole transformation. The Chern band of interest is below and above the Fermi energy (dotted line) in (a) and (b), respectively.}
\label{fig_reply1}
\end{figure}

We emphasize a technical detail in our model calculation: the calculation is performed in the “hole” basis. In this hole basis, the sum rule acts as a measure of the quantum geometry of the single Chern band below the Fermi energy at hole filling $\nu=1$ [Fig.~\ref{fig_reply1} (a)]. By performing a particle hole transformation back to the physical electron basis, the single Chern band becomes above the Fermi energy, as shown in Fig.~\ref{fig_reply1} (b).

Therefore, the sum rule with an integral of frequency limited to below $200$ meV measures the quantum geometry of the single Chern band right above the Fermi energy in the physical electron basis at hole filling $\nu=1$ [Fig.~\ref{fig_reply1} (b)]. We note that this Chern band is of particular interest, as it acts as the parent band to host fractional Chern insulators at fractional hole fillings.

To measure the quantum geometry of all the states below the Fermi energy in the physical electron basis (including those that are not captured by the continuum model), it is necessary to include optical absorption up to high energies in the sum rule, for example, the exciton states near 1 eV.

\subsection{Origin of the dominance of a single exciton}
Our calculation shows that a single exciton state can dominate the optical sum rule.
Here we show that the dominance of a single exciton is a quantum-mechanical phenomenon of coherent superposition in the exciton wavefunction.  We rewrite the exciton wavefunction as
\begin{equation}
|\chi\rangle=\sum_{\bm k,\lambda \ge 2}z_{\bm k,\lambda} (\chi) f_{\bm k\lambda+}^{\dagger} f_{\bm k1+}|G\rangle,
\end{equation}
where $z_{\bm k,\lambda} (\chi)$ is the envelope function obeying the normalization condition $\sum_{\bm k,\lambda \ge 2}|z_{\bm k,\lambda} (\chi)|^2=1$. Here $\lambda$ is the band index. The optical matrix element between $|\chi\rangle$ and the ground state $|G\rangle$ can be expressed as 
\begin{equation}
V_{\chi G}^{\alpha}=\langle\chi|\hat{v}^\alpha| G\rangle=\sum_{\bm k, \lambda\geq2}\left[z_{\bm k,\lambda}(\chi)\right]^*  v_{+,\lambda 1}^{\alpha}(\bm k),
\end{equation}
where $v_{+,\lambda 1}^{\alpha}(\bm k)$ is the optical matrix element in the basis of Bloch wavefunctions. We focus on the negative first moment,
\begin{equation}
W^{(1)}_{xx}=\frac{e^2}{\hbar}\frac{\pi}{\mathcal{A}}\sum_{\chi}\frac{|V_{\chi G}^{x}|^2}{\mathcal{E}^2_{\chi}},    
\end{equation}
where $\mathcal{A}$ is the system area. We note that $W^{(1)}_{xx}$ is an intensive quantity that does not scale with the system size. We now compare the case without and with exciton formation.

\begin{figure}[t]
\centering
\includegraphics[width=0.65\textwidth,trim=0 0 0 0,clip]{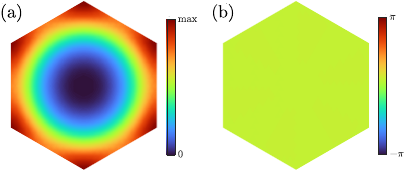}
\caption{(a) Amplitude and (b) phase of $\left[z_{\bm k,2}\right]^* v_{+,2 1}^{+}(\bm k)$ in the Brillouin zone for the dominant exciton state.}
\label{fig_reply4}
\end{figure}

\begin{figure}[b]
\centering
\includegraphics[width=0.75\textwidth,trim=0 0 0 0,clip]{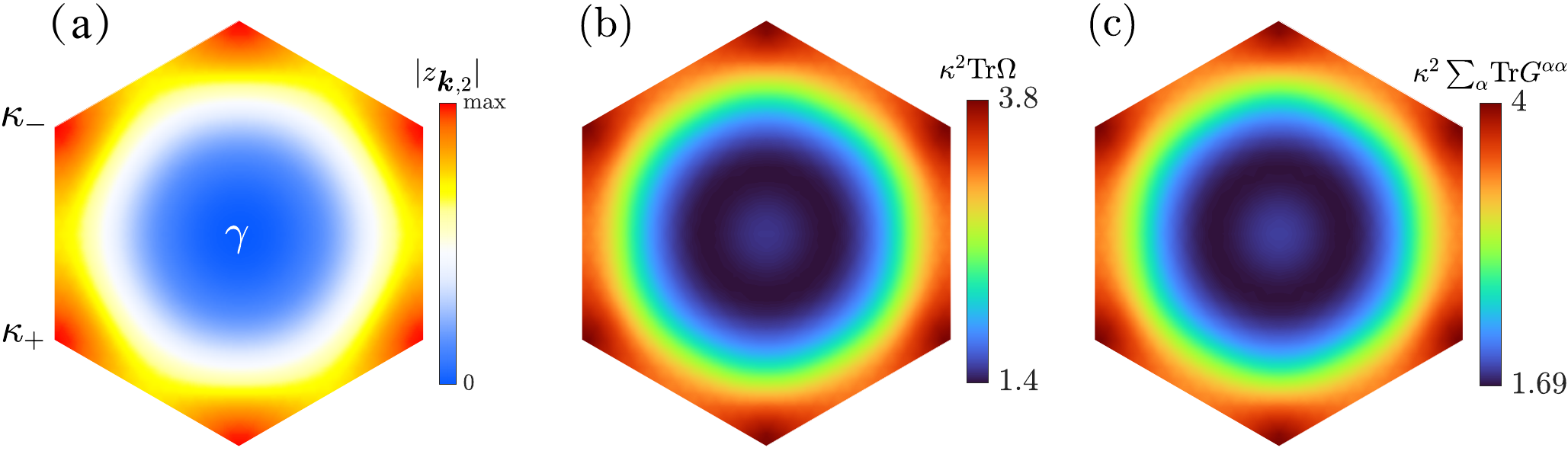}
\caption{(a) $|z_{\bm k,2}|$ in the moir\'e Brillouin zone for the dominant exciton state. (b) Berry curvature in the moir\'e Brillouin zone. (c) Trace of quantum metric in the moir\'e Brillouin zone. $\kappa$ is $4\pi/(3a_M)$. The calculations are performed for the CI state at $\theta=3.5^\circ$.}
\label{fig_reply5}
\end{figure}

(1) In the absence of exciton formation, $|\chi\rangle$ can also describe a particle-hole excitation but with the function $z_{\bm k,\lambda} (\chi)$ being finite only for a particular momentum $\bm k_0$ and band index $\lambda_0$, $z_{\bm k,\lambda} (\chi)=\delta_{\bm k,\bm k_0}\delta_{\lambda,\lambda_0}$. Then $|V_{\chi G}^{x}|^2$ does not scale with the system size; the contribution of a single excited state (without exciton formation) to $W^{(1)}_{xx}$, given by $\frac{1}{\mathcal{A}}\frac{|V_{\chi G}^{x}|^2}{\mathcal{E}^2_{\chi}}$, is vanishingly small due to the $\mathcal{A}$ factor in the denominator.

(2) The formation of excitons can drastically change the physics. We take the dominant exciton state in our study as an example. For this state, the envelope function $z_{\bm k,\lambda}$ is dominated by excitation to the second band (i.e., $\lambda=2$). As shown in Fig.~4(a) of the main text,$z_{\bm k,2}$ spreads over the momentum space instead of being concentrated at singular momentum points. With normalization $\sum_{\bm k,\lambda \ge 2}|z_{\bm k,\lambda} (\chi)|^2=1$, we have the scaling relation $z_{\bm k,2}\propto\frac{1}{\sqrt{N_k}}$, where $N_k$ is the number of momentum points in the Brillouin zone and proportional to the system area.

This dominant exciton state is chiral, and therefore, $|V_{\chi G}^{x}|^2=\frac{1}{2}|V_{\chi G}^{+}|^2=\frac{1}{2}|\sum_{\bm k, \lambda\geq2}\left[z_{\bm k,\lambda}(\chi)\right]^* v_{+,\lambda 1}^{+}(\bm k)|^2\approx \frac{1}{2}|\sum_{\bm k}\left[z_{\bm k,2}(\chi)\right]^*  v_{+,2 1}^{+}(\bm k)|^2$. A remarkable property comes from coherent superposition as reflected by the nearly uniform phase of $\left[z_{\bm k,2}(\chi)\right]^* v_{+,2 1}^{+}(\bm k)$ over the Brillouin zone, which is illustrated in Fig.~\ref{fig_reply4}. Therefore, there are $N_k$ terms (each scales as $\frac{1}{\sqrt{N_k}}$) that contribute coherently to the optical matrix element $V_{\chi G}^{x}$. Due to this coherence, $V_{\chi G}^{x}$ scales as $\sqrt{N_k}$ and $|V_{\chi G}^{x}|^2$ scales with the system area $\mathcal{A}$. This leads to the conclusion that a single exciton state can make a finite relative contribution to the sum rule. The dominance is further induced by the energy denominator $\mathcal{E}^2_{\chi}$, as the exciton has a low energy compared to higher-energy excited states.

\begin{figure}[t]
\centering
\includegraphics[width=0.75\textwidth,trim=0 0 0 0,clip]{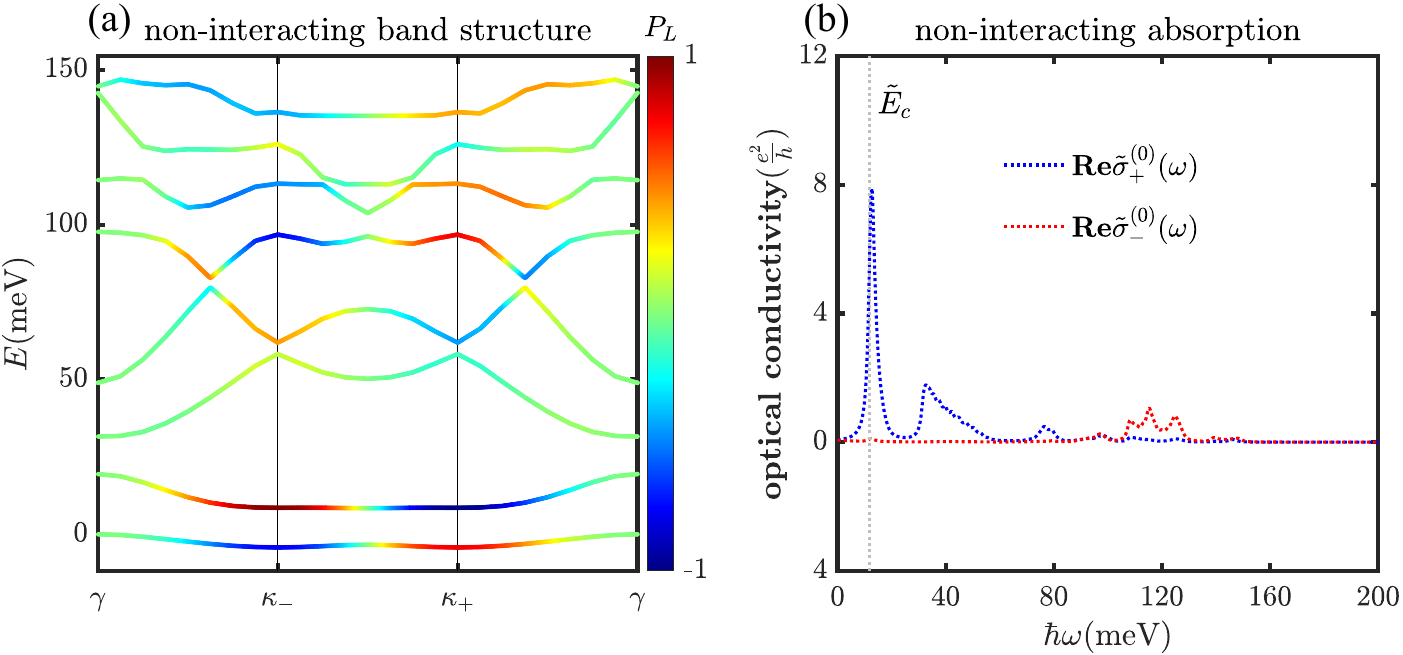}
\caption{(a) Band structure at $\theta=3.5^\circ$ and $\tau=+$ valley calculated using non-interacting single-valley model. The color represents the layer polarization $P_L$ with $+1$ ($-1$) indicating the bottom (top) layer. (b) $\text{Re}\tilde{\sigma}^{(0)}_{\pm}(\omega)$ calculated using the band structure in (a). The vertical dotted line marks the minimum direct gap $\tilde{E_c}$ between the first and second bands in (a).} 
\label{fig_reply3}
\end{figure}

We find that the envelope function $|z_{\bm k,2}|$ has a similar momentum dependence as the quantum geometric quantities. In Fig.~\ref{fig_reply5}, we plot the momentum-space distribution of quantum geometry for the Chern band under study, including the Berry curvature and the trace of the quantum metric. They have similar patterns with the envelope function $z_{\bm k,2}$ of the dominant exciton state. This provides an intuitive picture on why a single exciton state can largely probe the quantum geometry of the ground state. Note that this observation is based on the specific model of $t$MoTe$_2$. Nevertheless, there is a reason behind this behavior, and therefore, this relation could be general. Both the excitonic envelope function and the quantum geometric quantities have maxima at the Brillouin zone corners, where the gap in the electronic band structure is minimum. For the excitonic state, the minimum gap between occupied and unoccupied bands is favorable for electron-hole transitions, explaining why the excitonic envelope function is maximum there. The quantum geometric quantities are generally also large near the gap minimum, where band mixing effects are strong.

\subsection{Kohn’s theorem}
A single collective mode dominating the sum rule is known, for example, in the quantum Hall effects. Here are two famous examples. (1) Kohn’s theorem analytically shows that the optical response of quantum Hall states in Landau levels only occurs at a single resonance, i.e., the cyclotron resonance \cite{CyclotronKohn1961}. (2) The magneto-roton collective excitation spectrum in the fractional quantum Hall effect is well captured by single-mode approximation, as shown in the work~\cite{GirvinMagneto1986}.

To clearly demonstrate Kohn’s theorem, we consider optical absorption of Landau levels at filling factor $\nu=1$, i.e., one electron per magnetic flux. For simplicity, we do not consider the spin degree of freedom, assuming full spin polarization. The single-particle Hamiltonian in the presence of an out-of-plane magnetic field $B$ is
\begin{equation}
\mathcal{H}=\frac{p^2_x}{2m}+\frac{(p_y+eBx)^2}{2m},
\end{equation}
where Landau gauge is used. The single particle wavefunction is $\psi_{n,k}(\bm r)=\frac{1}{\sqrt{L_y}}e^{i k y}u_n(x+\frac{\hbar k}{eB})$, where $n$ is the Landau level index ($n=0,1,2\cdots$), $k$ is the momentum along $y$ direction, and $u_n(x)$ is the wavefunction of the quantum harmonic oscillator in one dimension. The velocity operator is $v_x=i[\mathcal{H},x]=\frac{\hbar p_x}{m}$. By using ladder operators, it can be shown that $v_x \psi_{n,k}(\bm r)=\hbar \sqrt{\frac{\hbar \omega_c}{2m}}(\sqrt{n+1}\psi_{n+1,k}(\bm r)+\sqrt{n}\psi_{n-1,k}(\bm r))$, where $\omega_c=\frac{eB}{m}$ is the cyclotron frequency.

We now turn to the second-quantized language for the many-body problem. We use $c^\dagger_{n,k}$ and $c_{n,k}$ for the creation and annihilation operators for the single-particle state $\psi_{n,k}(\bm r)$. The velocity operator in the second-quantized form is expressed as 
\begin{equation}
\hat{v}_x=\sum_{n,k}\hbar\sqrt{\frac{\hbar \omega_c}{2m}}(\sqrt{n+1}c^\dagger_{n+1,k}c_{n,k}+h.c.)
\end{equation}
At $\nu=1$, the zeroth Landau level is fully occupied in the ground state, $|\Psi_0\rangle=\prod_{k}c^\dagger_{0,k}|vac\rangle$, where $|vac\rangle$ is the vacuum state without any particle. In the absence of electron interaction, the optically active states on top of $|\Psi_0\rangle$ are $c^\dagger_{1,k} c_{0,k} |\Psi_0\rangle$, which is massively degenerate since $k$ can take any allowed values. However, in the presence of electron interaction, the massive degeneracy of the excited states are broken, leading to a single many-body excitonic state $|\Psi_1\rangle$ with the energy $\hbar \omega_c$ that is responsible for the optical excitation. As constructed explicitly in Ref.~\cite{CyclotronKohn1961}, $|\Psi_1\rangle$ is a coherent linear combination of $c^\dagger_{1,k} c_{0,k} |\Psi_0\rangle$,
\begin{equation}
|\Psi_1\rangle=\frac{1}{\sqrt{N}}\sum_{k} c^\dagger_{1,k} c_{0,k} |\Psi_0\rangle,
\end{equation}
where $N$ is the number of states included in the summation so that $|\Psi_1\rangle$ is normalized. The expectation value of the particle number operator in the state $|\Psi_1\rangle$ is one electron and one hole. The optical matrix element is, 
\begin{equation}
\langle \Psi_1 | \hat{v}_x | \Psi_0 \rangle=\frac{1}{\sqrt{N}}\hbar\sqrt{\frac{\hbar \omega_c}{2m}}
\sum_{k,k'}\langle \Psi_0 | c^\dagger_{0,k} c_{1,k} c^\dagger_{1,k'} c_{0,k'} | \Psi_0 \rangle =\frac{1}{\sqrt{N}} \hbar \sqrt{\frac{\hbar \omega_c}{2m}} \sum_{k,k'}\delta_{k, k'}=\hbar \sqrt{\frac{\hbar \omega_c}{2m}} \sqrt{N}.
\end{equation}
The response function $\text{Re}\sigma_{xx}(\omega)$ is given by
\begin{equation}
\text{Re}\sigma_{xx}(\omega>0)=\frac{e^2}{\hbar}\frac{\pi}{\mathcal{A}}\frac{|\langle \Psi_1 | \hat{v}_x | \Psi_0 \rangle|^2}{\hbar\omega_c} \delta(\hbar\omega-\hbar\omega_c)=n_0 \frac{\pi e^2}{2m} \delta(\omega-\omega_c)
\end{equation}
where $n_0=\frac{N}{A}=\frac{eB}{h}$ is the electron density at $\nu=1$. This response function satisfies both the $f$-sum rule and the generalized sum rule of the first negative moment,
\begin{equation}
\int_0^{+\infty} d\omega  \text{Re}\sigma_{xx}(\omega) =\frac{\pi n_0 e^2}{2m},  \quad
\int_0^{+\infty} d\omega  \frac{\text{Re}\sigma_{xx}(\omega)}{\omega} =\frac{\pi e^2}{2m}\frac{n_0}{\omega_c}=\frac{\pi}{2}\frac{e^2}{h}.
\end{equation}
Here the second equation relates the first negative moment of $\text{Re}\sigma_{xx}(\omega)$ to the quantum weight $K$ of the lowest Landau level, which is exactly 1.

Therefore, the single many-body state $|\Psi_1\rangle$ fully dominates the optical sum rules. Kohn’s theorem is more general than the above demonstration, as it applies to other filling factors as well. We use Landau gauge wavefunctions in the above derivation, which can be alternatively formulated in the two-dimensional magnetic Brillouin zone using magnetic Bloch states with  two-dimensional wavevectors.

We note that some of the important physical ingredients in Kohn’s theorem is shared in $t$MoTe$_2$. (1) In $t$MoTe$_2$, the Chern band under study has a narrow bandwidth, mimicking the Landau level physics both in terms of bandwidth and the Chern number.  (2) In Kohn’s theorem, both the envelope function in the excitonic state and the quantum geometric quantities are independent of momenta (uniform in the Brillouin zone). In $t$MoTe$_2$, the excitonic envelope function, although not uniform in the Brillouin zone, has a similar momentum dependence as the quantum geometric quantities.

\section{Results for non-interacting system}
In Fig.~1(b) of the main text, $\text{Re}\sigma^{(0)}_{\pm}$ denotes the absorption on top of the Hartree-Fock band structure at $\nu=1$ without including the electron-hole interaction in the excited states, but the Hartree-Fock electronic band structure does include interaction effects.

For comparison, we also plot the non-interacting “single-valley” band structure in Fig.~\ref{fig_reply3}(a) for valley $\tau=+$. Compared to the Hartree-Fock band structure in Fig.~1(a) of the main text, the non-interacting band structure has a smaller gap separating the first and second bands. We use $\text{Re}\tilde{\sigma}^{(0)}_{\pm}$ to denote the single-particle absorption for optical transition between the first and other bands of the non-interacting “single-valley” band structure. The spectrum of $\text{Re}\tilde{\sigma}^{(0)}_{\pm}$, as shown in Fig.~\ref{fig_reply3}(b), has a similar circular dichroism as the spectrum of $\text{Re}\sigma^{(0)}_{\pm}$ shown in the main text. Because no interaction effects are included in the calculation of $\text{Re}\tilde{\sigma}^{(0)}_{\pm}$, there are no excitonic resonances. The strong peak in $\text{Re}\tilde{\sigma}^{(0)}_{\pm}$ right above the gap results from the narrow bandwidths of the first and second bands, which leads to a large joint density of states.

\section{Control of circular dichroism by valley polarization}
\begin{figure}[t]
\centering
\includegraphics[width=0.4\textwidth,trim=0 0 0 0,clip]{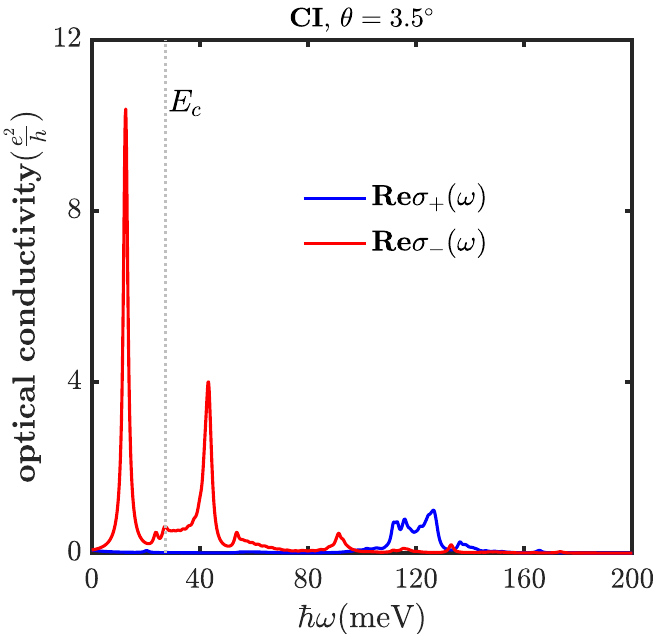}
\caption{$\text{Re}\sigma_{\pm}(\omega)$ for the CI state with holes polarized in $\tau=-$ valley. The vertical dotted line marks the minimum direct gap $E_c$.} 
\label{fig_reply2}
\end{figure}

In the Hartree-Fock calculation, the initial ansatz for the many-body state is taken to be holes fully doped to $\tau=+$ valley, which generates the valley polarized states after self-consistent iteration calculation, as presented in the main text. 
Equivalently, we can take another initial ansatz with holes fully doped to $\tau=-$ valley, which generates a CI with opposite valley polarization and Chern number after the self-consistent calculation. The chirality of the spectrum is also flipped, as shown in Fig.~\ref{fig_reply2}. 
The two CI states with opposite valley polarizations are time-reversal partners, but each one individually breaks time-reversal symmetry spontaneously. Experimentally, the two CI states can be selected by applying a training out-of-plane magnetic field.

\end{document}